\documentclass[lettersize,journal]{IEEEtran}
\usepackage{amsmath,amsfonts}
\usepackage{algorithm}
\usepackage{algorithmic}

\usepackage{array}
\usepackage{booktabs}
\usepackage{multirow}
\usepackage{adjustbox}
\usepackage{longtable}
\usepackage{tabularx}

\usepackage{graphicx}
\usepackage[caption=false,font=normalsize,labelfont=sf,textfont=sf]{subfig}

\usepackage{textcomp}
\usepackage{stfloats}
\usepackage{verbatim}
\usepackage{caption}

\usepackage{url}
\usepackage{cite}
\usepackage[colorlinks=true, linkcolor=blue, citecolor=blue, urlcolor=blue]{hyperref}

\usepackage{color,soul} 
\sethlcolor{yellow} 

\makeatother

\def\BibTeX{{\rm B\kern-.05em{\sc i\kern-.025em b}\kern-.08em
    T\kern-.1667em\lower.7ex\hbox{E}\kern-.125emX}}
\begin{document}

\title{A Survey of Smart Grid Emerging Use Cases and Relevant 5G and 6G Capabilities and Features}

\author{Manoj Kumar,~\IEEEmembership{Graduate Student Member,~IEEE,} Nishith D. Tripathi,~\IEEEmembership{Senior Member,~IEEE,} Jeffrey H. Reed, ~\IEEEmembership{Life Fellow,~IEEE}

\thanks{All authors are affiliated with Wireless@VT at The Bradley Department of Electrical and Computer Engineering, Virginia Tech, Blacksburg, VA, USA - 24061 (email: \{manoj.kumar, nishith, reedjh\}@vt.edu)}

\thanks{This work has been peer reviewed, accepted and published in IEEE Access. This work is licensed under a Creative Commons Attribution 4.0 License. For more information, see https://creativecommons.org/licenses/by/4.0/ \\ DOI: 10.1109/ACCESS.2026.3667671}}
\maketitle

\begin{abstract}
The growing complexity of modern energy systems has led to the adoption of Smart Grid (SG) that use advanced communication technologies to facilitate efficient, reliable, secure, and sustainable energy operation and management. Unlike existing surveys that often treat grid and communication domains separately, this work rigorously quantifies service requirements for high-complexity emerging scenarios. It provides a comprehensive overview of SG architecture that integrates digital communication infrastructure with distributed energy resources (DERs), microgrids, energy storage systems, and cybersecurity frameworks. Furthermore, emerging SG use cases such as smart distributed voltage control, real-time fault detection and self-healing, smart and autonomous monitoring, and predictive maintenance are identified, and more importantly, service performance requirements associated with these use cases have been quantified. Additionally, key capabilities and emerging SG enablers of fifth-generation  (5G) and sixth-generation (6G) networks are described.  These capabilities and enablers include network slicing, edge computing, spectrum management, artificial intelligence (AI) driven optimization, digital twins, and Open-Radio Access Network (O-RAN). Finally, the paper discusses open challenges and future research directions for designing scalable, intelligent, and secure next-generation SG systems.
\end{abstract}

\begin{IEEEkeywords}
5G, 6G, digital twin, edge computing, emerging use cases, network slicing, O-RAN, predictive maintenance, Smart Grid.
\end{IEEEkeywords}

\section{Introduction}
\label{sec:introduction}
\IEEEPARstart{C}{ellular} networks play a vital role in providing wireless connectivity for mobile phones and Internet of Things (IoT) devices. The first generation (1G) primarily focused on analog voice communication\cite{b1}. The second generation (2G) introduced digital technology, which improved voice quality and added Short Message Service (SMS) \cite{b1}. The third generation (3G) advanced data and multimedia capabilities, facilitating features like video conferencing \cite{b1, b2}. The fourth generation (4G) significantly enhanced data transmission speeds with Long-Term-Evolution (LTE) technology, allowing for high-definition video streaming and mobile gaming \cite{b1}. The 5G as defined by 3$^\text{rd}$ Generation Partnership Project (3GPP) in Release 15, represents a major milestone in cellular evolution. It is designed to meet the performance requirements established by the International Telecommunication Union (ITU) under International Mobile Telecommunications-2020 (IMT-2020). 5G is intended to facilitate a range of applications. It supports scenarios such as enhanced mobile broadband (eMBB), ultra-reliable low-latency communications (URLLC), and massive machine-type communications (mMTC) \cite{b3}. Key performance targets for 5G include peak data rates of up to 20 Gigabits per second (Gbps), network latencies as low as 1 milliseconds (ms), and the capacity to accommodate up to 1 million IoT devices per square kilometer (km$^\text{2}$) \cite{b3}. At its essence, 5G is built on the New Radio (NR) air interface and advanced network architectures, including innovations such as virtualization and network slicing \cite{b1}. 

Subsequent releases of 3GPP continue to enhance 5G’s capabilities. Release 16 introduces features such as NR unlicensed (NR-U), enhanced vehicle-to-everything (eV2X), and improved industrial IoT support. Release 17 brings forth non-terrestrial network (NTN), higher frequency bands, and further refinements in NR sidelink and Reduce Capability (RedCap) devices \cite{b3, b25}. Release 18 is expected to focus on improvements in non-public networks, unmanned aerial vehicle (UAV) connectivity, and media streaming. Release 19 introduces advancement in areas such as extended reality (XR), artificial intelligence/machine learning (AI/ML) enhancements, satellite architecture such as regenerative payload NTN, energy efficiency as a service, and Ambient IoT \cite{b3, b25}. It also includes IoT over NTN and support for tactile and immersive communication services through Integrated Sensing and Communication (ISAC). Further improvements are expected in XR and AI/ML for the Next Generation Radio Access Network (NG-RAN) air interface \cite{b3, b25}. Looking forward, the emergence of next-generation 6G technology is projected around the year 2030. The ITU goals have been defined for 6G \cite{b23, b25}. Its aim is to support immersive applications like high-fidelity holograms, multisensory communication, terahertz (THz) communication, and advanced artificial intelligence (AI) driven services \cite{b3, b23, b32}. 

The SG is a smart and intelligent infrastructure of an electricity network. It uses cutting-edge digital technologies, intelligent, and advanced communications techniques, to ensure the SG is more reliable, intelligent, secure, responsive, and efficient and improves the overall SG operations and usage experience \cite{b4, b18, b19, b22}. The SG marks a significant advancement from traditional centralized power grids to a more dynamic and decentralized energy network. It effectively integrates real-time monitoring, two-way communication, and DERs, establishing a new paradigm for energy management \cite{b4, b6}. In contrast to the traditional grid, the SG employs advanced communication infrastructure that provides automated control, predictive maintenance, early fault detection and improved reliability through self-healing capabilities \cite{b6, b7, b4}. 

Distributed generation, particularly from renewable sources such as solar and wind, supports sustainable energy objectives. While significantly enhancing resilience. Localized microgrids and virtual power plants (VPPs) \footnote{A VPP is a cluster of distributed generators such as solar panels, wind turbines, and other DERs that are managed by a central controller. It operates collectively to deliver power with efficiency and flexibility comparable to a conventional power plant. While enabling rapid response to grid fluctuations and market conditions \cite{b4}.} are crucial in stabilizing grid operation during peak demand periods or outages \cite{b4}. Key regulatory frameworks provided by organizations such as Institute of Electrical and Electronics Engineers (IEEE) and National Institute of Standards and Technology (NIST) establish essential interoperability and cybersecurity standards, ensuring secure and efficient delivery of energy \cite{b5}. Furthermore, consumer engagement is revolutionized through smart metering and demand response programs, leading to substantial improvements in energy efficiency and load management \cite{b4}. The SG serves as a resilient and adaptable foundation ready for future innovations, including AI-driven optimization and maintenance. Making it indispensable to address the energy challenges of the present and future \cite{b6}.

 Wireless communication is essential to delivering a flexible and cost-effective approach to managing the demands for energy and a variety of devices over large areas \cite{b7}. The key characteristics of the SG requires dependable communication networks to facilitate DERs, advanced metering infrastructure, and real-time monitoring. Technologies such as 5G and 6G provide the quality of service (QoS) such as coverage, high data rates, and low latency, ensuring rapid responses to grid events and providing quick fault detection and self-healing capabilities \cite{b7}. Wireless communication solutions reduce deployment and maintenance costs compared to traditional wired networks, making them more practical for areas where infrastructure may be limited, such as in remote or rural areas \cite{b7}. It also provides the scalability required to accommodate new technologies, including electric vehicles (EVs) and residential energy storage, which require robust communications to integrate effectively with the SG\cite{b8}. 
 
 Moreover, wireless technologies such as 5G/6G allow demand-response programs and variable pricing. It allows consumers to adjust energy usage based on real-time data. By supporting these functionalities, wireless communication boosts grid reliability and efficiency, which is critical for sustainable energy transition \cite{b7, b8, b9}. Recent studies have begun to explore specific intersections of 6G and smart energy systems. For instance, the authors \cite{b48} investigated AI-driven security architectures, while \cite{b49} examined sustainability and the role of 6G in reducing the carbon footprint. While these contributions are vital, they predominantly focus on cyber-defense and energy efficiency rather than the deterministic performance required for physical grid protection. 
 
 Unlike existing literature that treats grid and communication as separate or only partially integrated, this work quantifies the service requirements for complex SG use cases and evaluates the impact of Network Slicing, ISAC, AI-native networking, and O-RAN on next-generation energy systems.
The main contributions of this paper are as follows:
\begin{itemize}
\item We present a holistic SG architecture aligning the NIST conceptual model with DERs, microgrids, and cybersecurity frameworks to ensure grid reliability and flexibility.
\item We identify novel scenarios that include AI/ML-based Falling Conductor Detection and Proactive Climate Resilience, and quantitatively define their stringent performance benchmarks (e.g., availability, latency).
\item We conduct a comparative evaluation of 5G and 6G (IMT-2030) capabilities, specifically analyzing the transition to transformative features like ISAC and AI-native networking, to demonstrate their utility in securing and automating grid operations.
\item We map specific grid use cases to essential communication key performance indicators (KPIs), such as time synchronization and message size, to bridge the gap between theoretical specifications and practical utility.
\item We analyze enabling technologies often overlooked in surveys, specifically O-RAN, Network Slicing, and Digital Twins, to define a research agenda for intelligent energy systems.
\end{itemize}
The rest of this paper is organized as follows. Section \ref{sec:conprimer} provides a detailed overview of the architecture of SG that focuses on DERs and the factors that drive the adoption of SG. Section \ref{sec:wiredvswireless} compares wired and wireless communication within the SG, discussing aspects such as reliability, cost, security, speed, flexibility, and scalability.  Section \ref{sec:useserv} examines the emerging use cases of SG and summarizes their corresponding service requirements.  Section \ref{sec:capab} assesses the role of 5G and 6G technologies and considers potential challenges associated with the development of SG.  Section \ref{sec:challenges} presents the implementation plan of the 5G/6G-based SG by discussing the formation of a consortium. The development of scheduling and resource allocation strategies and customization of networks to meet the specific needs of the SG applications. Eventually, Section \ref{sec:smmry} summarizes the critical 5G and 6G specifications, such as sub-millisecond latency, integrated sensing, Digital Twin, and AI-native networking, essential for supporting future SG applications, and highlights current limitations and future research directions.

\section{SG : A Concise Primer}
\label{sec:conprimer}
The SG provides real-time surveillance and management of the generation, transmission, and distribution of electricity, facilitating two-way communication between consumers and service providers of electricity. This allows customers to generate and put energy back from renewable sources like solar and wind, helping better management of electricity usage and costs. The SG facilitates clean energy integration, improving operational resilience, sustainability, and overall effectiveness of electricity services \cite{b6, b19}. 

\begin{figure}[!t]
\centering
\includegraphics[width=3.5in]{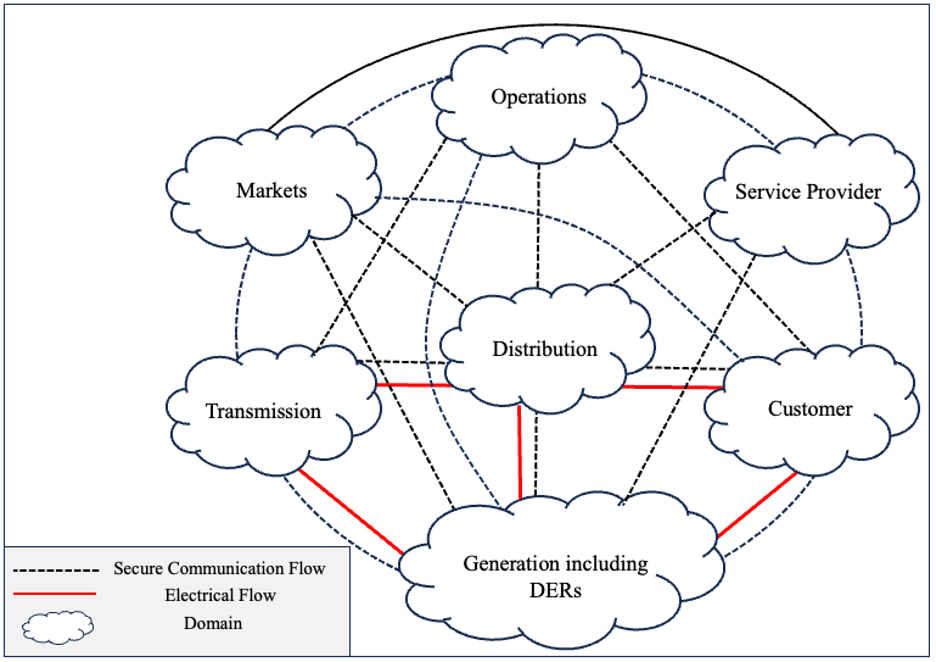}
\caption{The NIST Conceptual View of a SG\cite{b28}}
\label{fig:NISTConcViewSG}
\end{figure}

In this study, the system boundaries are defined by the NIST conceptual model, which includes utility-scale renewable plants within the Generation (including DERs) domain. This encompasses Concentrated Solar Power (CSP) fields as part of the broader class of renewable generation assets addressed by this survey. The conceptual view of the SG specified by NIST is shown in Fig. \ref{fig:NISTConcViewSG}. Fig. \ref{fig:NISTConcViewSG} illustrates the interconnected ecosystem of an SG, highlighting the interaction between main domains (cloud represents domain) such as generation including DERs, transmission, distribution, markets, operations, service providers, and customers \cite{b28}. Electricity flow, indicated by red lines, represents the bidirectional energy transfer among generation, transmission, distribution, and customer, promoting renewable energy integration and prosumer activities. Prosumers are the customers, having capabilities of both producing and consuming electricity. Meanwhile, secure communication flows (dashed lines) illustrate the exchange of real-time data among all the domains, facilitating advanced monitoring, automation, and decision-making. This interconnected structure supports SG efficiency, reliability, and sustainability by integrating renewables, optimizing demand response \footnote{Demand response refers to a set of strategies and technologies that enable electricity consumers to adjust their energy usage either manually or automatically in response to real-time signals such as price changes, peak demand events, or grid reliability conditions. It helps balance supply and demand, reduce strain on the grid, and lower operational costs.}, and boosting market and service innovations such as variable pricing, customized services, integration of renewables, and support for EVs \cite{b28}.

\begin{figure*}[!t]
\centering
\includegraphics[width=0.98\textwidth]{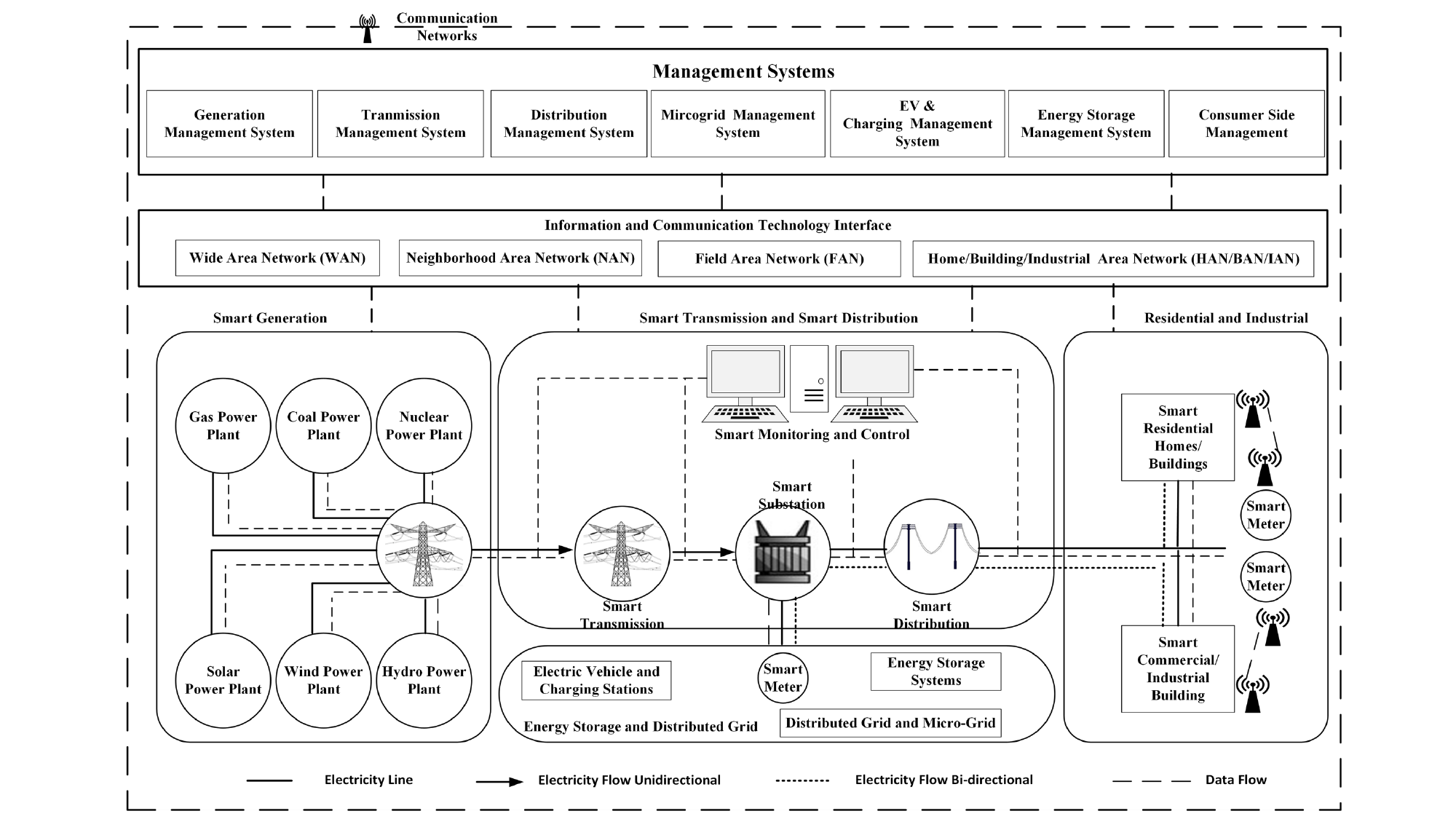}
\caption{Overall Architecture of SG\cite{b6, b28, b4, b9, b18, b19}}
\label{fig:sgarchitecture}
\end{figure*}

Fig. \ref{fig:sgarchitecture} shows the architecture of the SG system, which integrates generation, transmission, distribution, and consumers with advanced communication technologies, DERs, and real-time management systems to create an efficient and intelligent energy ecosystem \cite{b28}. The main elements of the architecture are management systems, Information and Communication Technology (ICT) interfaces, Smart Generation, Smart Transmission, Smart Distribution, Consumers (e.g., smart residential/homes/commercial /industrial), energy storage, distribution grid, microgrid, monitoring systems, smart meters, IoT devices (e.g., phasor measurement unit (PMU)) and communication networks. These systems and mechanisms enhance grid reliability, efficiency, sustainability, and active consumer participation. A detailed explanation of architecture is given below. 
\begin{itemize}
\item{\bf Management Systems:}
Management systems include Generation, Transmission, and Distribution Management Systems. They ensure optimal energy generation, transmission, and distribution. These are complemented by systems for microgrid management, EV and charging infrastructure management, energy storage, and consumer side management, providing efficient operation and control of diverse energy resources \cite{b6} \cite{b28} \cite{b4}.

\item{\bf ICT Interface:}
The SG’s ICT interface links management systems to field devices and consumers through a hierarchical network structure. Wide Area Networks (WAN) manage long-distance communication, while Neighborhood Area Networks (NANs), Field Area Networks (FANs), and Home/Building/Industrial Area Networks (HAN/BAN/IAN) connect localized devices, smart meters, and appliances, supporting seamless real-time monitoring and controls \cite{b4} \cite{b18} \cite{b19}.

\item{\bf Smart Generation System:}
The Smart Generation System incorporates both conventional power plants (such as gas, coal, and nuclear) and renewable energy sources (such as solar, wind, and hydro). This dual-generation approach ensures a stable and continuous energy supply while encouraging sustainability by integrating renewable resources. The inclusion of DERs like rooftop solar panels facilitates localized energy generation and the capability of putting energy back by consumers to the grid, reducing dependency on centralized power plants and enhancing grid flexibility. As renewable sources like wind and solar are variable, the integration of DERs and energy storage and management systems ensures grid stability and ecofriendly power generation \cite{b4} \cite{b18} \cite{b19}.

\item{\bf Smart Transmission:}
The smart transmission system is responsible for delivering electricity from generation plants to distribution systems through high-voltage transmission lines, utilizing the existing power transmission infrastructure. It incorporates advanced technologies such as smart substations and smart monitoring systems, which are equipped with digital information, automation, communication, and self-healing capabilities. These features support real-time fault detection, minimize transmission losses, and enable more efficient energy flow. Additionally, smart transmission enhances reliability by dynamically responding to changes in demand and generation patterns. Transmission monitoring systems are essential for improving system efficiency and facilitating the integration of renewable energy resources \cite{b4, b6, b19}.

\item{\bf Smart Distribution:}
The smart distribution system delivers electricity from transmission systems to end users such as residential, commercial, and industrial consumers. This system supports two-way energy flows (dashed-dotted line), allowing surplus energy from DERs (e.g. residential solar panels) and energy storage systems (e.g., inverters) to be put back into the grid. Key features include smart monitoring control systems such as smart and intelligent meters, IoT devices, and communication networks, which allow real-time monitoring, efficient distribution of energy, and improved grid reliability \cite{b4, b6, b18}. 

\item{\bf Residential and Industrial Consumers:}
Residential and industrial consumers are equipped with smart meters and IoT devices, which enable real-time monitoring of electricity consumption and energy management. Consumers, referred to as prosumers, can also generate energy through renewables such as rooftop solar panels and sell surplus electricity back to the grid. The SG allows the active participation of consumers through smart meters, IoT devices, and demand response programs. This promotes energy efficiency, cost savings, and grid stability \cite{b4, b18, b6, b9}.

\item{\bf Energy Storage and Distributed Grid:}
The energy storage and management systems (e.g., batteries), distributed grids, and microgrids significantly enhance the SG’s resilience, flexibility, and decentralization. Distributed grids integrate geographically dispersed DERs into the central grid to improve efficiency and reduces transmission losses, while microgrids operate as localized energy networks capable of functioning independently during grid outages.  These systems store extra energy during periods of low demand and release it during peak demand hours or outages, ensuring reliability and load balancing. Additionally, the integration of EVs Charging Stations and Vehicle to Grid (V2G) technology allows EVs to act as mobile storage units, contributing to grid flexibility and stability \cite{b4, b6, b19}.

\item{\bf Data Flow and Electricity Flow:}
In Fig. \ref{fig:sgarchitecture}, the solid black lines with unidirectional arrows show a unidirectional flow of electricity for traditional power delivery. The dashed-dotted black line represents a bi-directional flow of electricity for the integration of DERs and energy storage systems. Doubled-dashed black lines show the exchange of real-time data across all grid elements. The data flow is essential for dynamic decision-making and optimization of operations \cite{b6}. The dashed lines show that the entire SG infrastructure is covered by communication systems.

\item{\bf Cybersecurity Considerations:}
As the SG integrates millions of IoT devices, smart meters, and DERs, cybersecurity becomes a critical priority. Risks such as ransomware attacks, data manipulation, and unauthorized system access can disrupt operations and compromise grid stability. To address these risks, implementing zero-trust architectures, cryptographic protocols, and secure communication standards has been recommended to protect data flows and operational infrastructures \cite{b6}.
\end{itemize}

In conclusion, the SG system integrates renewable energy, energy storage, advanced communication networks, and consumer participation to create a sustainable and efficient energy ecosystem. It provides real-time data monitoring and automation to address the challenges of modern energy demands while promoting a cleaner and more decentralized energy system. 

\begin{figure*}[!t]
\centering
\includegraphics[width=\textwidth]{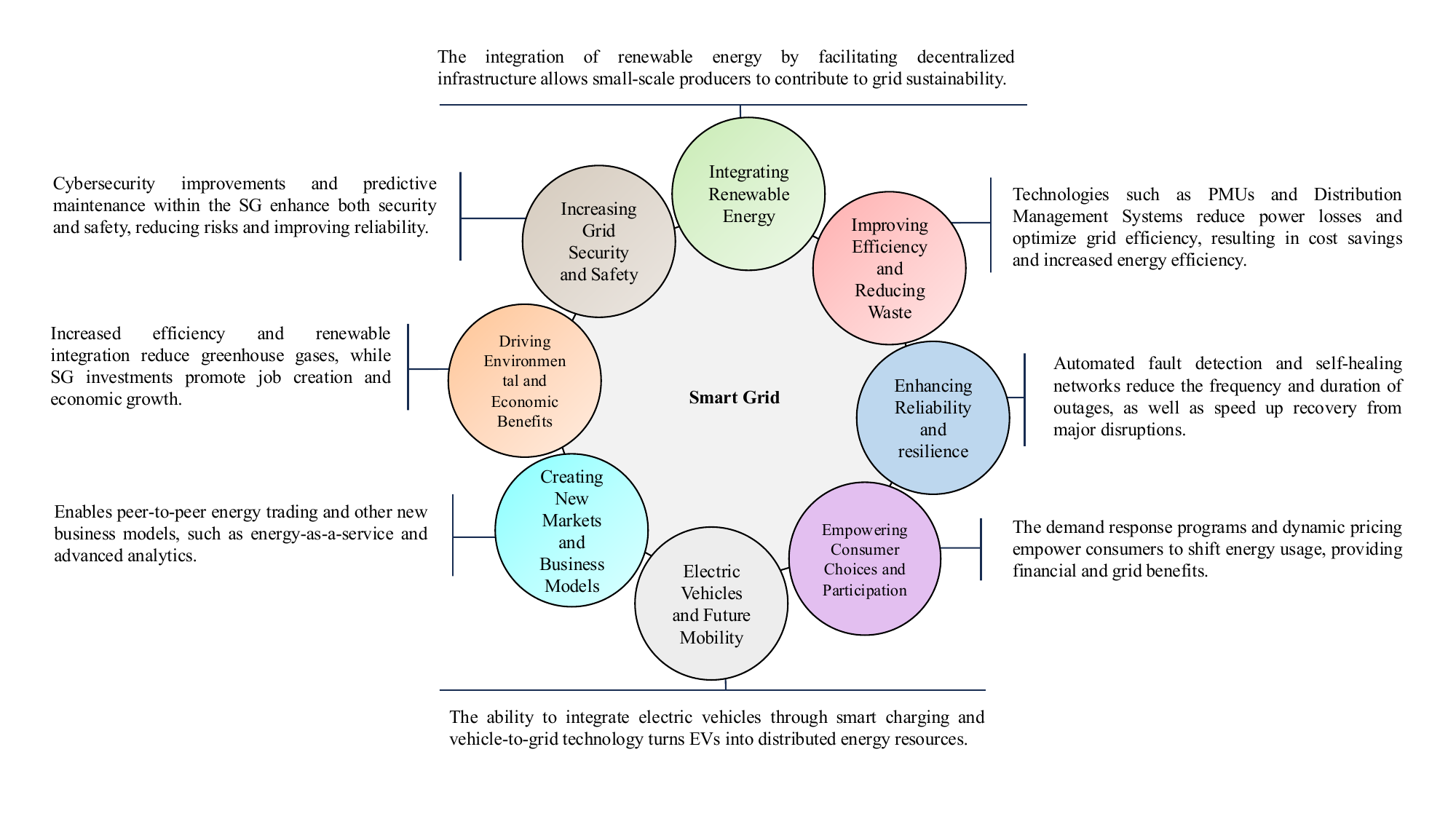}\caption{A SG as a Platform for New Possibilities\cite{b6, b9, b18, b19, b28}}
\label{fig:sgpltnewposs}
\end{figure*}

The SG is becoming increasingly popular for several reasons, such as emphasizing renewable energy integration, energy efficiency improvements, enhanced reliability, and cybersecurity. It supports consumer participation, promotes economic benefits, fosters new business models, and inspires EV integration, transforming traditional grids into efficient, sustainable, and consumer-driven energy systems as shown in Fig. \ref{fig:sgpltnewposs}. Its adoption supports environmental sustainability, fosters economic growth, and promotes innovative energy markets, making it a cornerstone of the energy transition \cite{b6, b9, b28}.

The key differences between traditional grid and SG are summarized in Table \ref{tab:tgvssg}. However, realizing this decentralized, highly responsive SG architecture is impossible without a robust underlying data exchange mechanism, which brings us to the critical evaluation of grid communication media.

\begin{table}[!t]
\caption{Traditional Grid vs SG\label{tab:tgvssg} \cite{b4, b6, b9, b18, b19}}
\centering
\begin{tabular}{|p{55pt}|p{75pt}|p{90pt}|}
\hline
\bf{Aspects} & \bf{Traditional Grid} & \bf{SG} \\
\hline
Communication and Data Flow & One-way, limited communication, lacks real-time monitoring \cite{b4, b6, b18, b19} & Two-way digital communication, allows real-time data exchange, monitoring and automated control \cite{b4,b6, b19} \\
\hline
Operational Structure & Centralized generation and control, inflexible and slow to respond to demand changes \cite{b9, b19} & Decentralized with distributed generation (e.g., renewables), adaptive and responsive to demand changes \cite{b18, b19, b9} \\
\hline
Reliability and Resilience & Prone to outages and cascading failures, primarily manual restoration \cite{b4, b19} & Self-healing capabilities with automated fault detection and isolation \cite{b4, b18, b19}\\
\hline
Environmental Impact & High reliance on fossil fuels, limited renewable integration \cite{b4, b9} & Integrates renewable sources effectively, reduces carbon emissions \cite{b19, b9, b4}\\
\hline
Consumer Interaction & Minimal consumer engagement, limited data on usage, static pricing \cite{b9, b4, b19} & Active consumer role with smart meters, real-time usage data and demand response capabilities \cite{b19, b4, b9} \\
\hline
Security and Cyber Resilience & Basic physical security, limited cyber defenses \cite{b6, b4, b9} & Advanced cybersecurity with AI-driven systems, secure communication protocols \cite{b6, b4, b9, b18} \\
\hline
Economic Efficiency & High operational and infrastructure costs due to inefficiency and centralization \cite{b19, b4, b9} & Optimized resource allocation, reduces infrastructure and operational costs \cite{b4, b6, b9, b18} \\
\hline
\end{tabular}
\end{table}

\section{Wired and Wireless Communications}
\label{sec:wiredvswireless}
In an SG implementation, the communications infrastructure is a crucial part, allowing efficient data transmission for real-time monitoring, automated control, and optimization across grid elements. The primary communication approaches can be classified into wired communication and wireless communication, each offering unique benefits and challenges for specific SG applications. A significant number of works have been conducted on wired and wireless communication in SG.

\subsection{Wired Communication}
In an SG, wired communications, such as Power Line Communication (PLC), Ethernet, and fiber optics, are recognized for their high reliability, low latency, and enhanced security, which are essential for mission-critical grid operations. These technologies support the stability required for applications like Wide Area Monitoring Systems (WAMS), Supervisory Control and Data Acquisition (SCADA), and Advanced Metering Infrastructure, where uninterrupted data flow is crucial \cite{b10}. PLC, for instance, is advantageous in leveraging existing power lines, thus minimizing additional infrastructure costs in established grid environments. However, wired communication faces limitations, particularly in terms of installation costs and scalability. While PLC provides stable, low latency connection ideal for real-time applications such as fault detection, its rigid infrastructure requirements can impede deployment in geographically expansive or rapidly evolving grid segments \cite{b17}. These studies collectively suggest that wired communication is best suited for high-reliability applications within SG, where its inherent stability and security provide significant advantages despite cost constraints.

\subsection{Wireless Communication}
The wireless communications technologies of modern SG, including Wi-Fi, ZigBee, Bluetooth, and cellular networks (e.g., 5G and future 6G networks) are valuable due to their flexibility, scalability, and ease of deployment of SG, particularly in remote or distributed grid areas. The authors in \cite{b13} highlight that wireless systems offer adaptable and cost-effective solutions, permitting communication in areas where laying physical cables is impractical or cost-prohibitive. However, wireless systems are susceptible to interference, security, vulnerabilities, and variable performance under high data loads. Hence, these issues need to be addressed while using a wireless system.  

The emerging wireless technologies, such as 5G can support the growing data demands, reliability, security, faster data rate, and massive connectivity requirements for SG. Particularly for applications like DERs management and real-time monitoring in urban areas \cite{b15}. Wireless communication should be strategically implemented in non-critical applications, such as demand response, remote monitoring, and customer-end metering, where flexibility and scalability are prioritized over ultra-low latency and high security \cite{b13, b14}. 

While wireless systems face challenges in security, reliability, and susceptibility to interference, especially in data intensive and low-latency applications, emerging technologies such as 5G/6G offers significant benefits such as ubiquitous coverage, reduced latency, reliability, security, support for a massive number of devices, high data speed, sensing capabilities, positioning, AI capabilities, interoperability, and sustainability, making them increasingly viable for adaptive SG applications.

\subsection{Comparison between wired and wireless communications}
A comprehensive comparison of these communications techniques helps to identify their strengths, limitations, and ideal use cases within the SG. Table \ref{tab:wiredvswireless} summarizes key aspects of wired and wireless networks in the context of SG applications, highlighting factors like reliability, data rate, coverage, and installation costs \cite{b10, b11, b12, b13, b14, b15, b16, b17}.  

Beyond traditional cryptographic methods, the forthcoming 5G-Advanced and 6G systems are anticipated to utilize Physical Layer Security (PLS) techniques. PLS capitalizes on the inherent randomness of the wireless channel to maintain confidentiality. Examples of these techniques include channel-based secret key generation among grid devices, transmission methods enhanced with artificial noise to interfere with snoopers, and the implementation of reconfigurable intelligent surfaces (RIS) or holographic metasurfaces to optimize propagation conditions for legitimate receivers. By integrating these PLS mechanisms with higher-layer security protocols, the systems can partially mitigate the vulnerability of wireless SG links to eavesdropping and jamming, offering an improvement over legacy cellular generation\cite{b43}. While these advanced wireless mechanisms provide the foundation for a secure and scalable grid, their true value is realized through the complex, data-intensive applications they enable. Consequently, understanding the specific service requirements of these emerging use cases is critical.

\begin{table}[!t]
\caption{Comparison of Wired and Wireless Communication \cite{b10, b11, b12, b13, b14, b15, b16, b17} \label{tab:wiredvswireless}}
\centering
\begin{tabular}{|p{40pt}|p{75pt}|p{90pt}|}
\hline
\bf{Aspect} & \bf{Wired} & \bf{Wireless} \\
\hline
Definition & Data transmission via physical media like cables \cite{b10, b14} & Data transmission using radio signals \cite{b14, b15} \\
\hline
Technologies & PLC, Ethernet, Fiber Optics \cite{b10, b12, b16} & Wi-Fi, ZigBee, Bluetooth, 4G/5G, Long Range Wide Area Network (LoRaWAN) \cite{b12, b13, b15} \\
\hline
Reliability & High reliability with low susceptibility to interference \cite{b10, b12, b14} & Moderate reliability; vulnerable to interference \cite{b11, b13, b14, b15}\\
\hline
Latency & Low latency, suitable for real-time applications \cite{b10, b17} & URLLC latency $\leq$ 1 ms; may experience higher latency depending on network conditions \cite{b11, b13, b14, b15}\\
\hline
Installation Cost & High due to infrastructure requirements (cabling, trenching) \cite{b10, b12, b14} & Lower; requires minimal physical infrastructure \cite{b11, b12, b14, b15}\\
\hline
Flexibility & Limited; physical cables restrict expansion and reconfiguration \cite{b10}\cite{b12} & High; allows for easy network expansion and reconfiguration \cite{b13, b14, b15, b17}\\
\hline
Maintenance & Generally low, but susceptible to physical damage over time \cite{b10, b12}& Moderate; ongoing need for monitoring signal strength and interference \cite{b15, b17}\\
\hline
Security & Generally, more secure due to physical containment \cite{b10, b17} & Traditionally more vulnerable, but evolving with Physical Layer Security (PLS) and robust encryption in 5G/6G\cite{b12, b15, b17, b43}\\
\hline
Data Transfer Speed & Typically high, depending on cable type (e.g., fiber optic) \cite{b10, b14}& Varies; lower for some wireless protocols (e.g., ZigBee) \cite{b13, b14, b15}\\
\hline
Coverage Range & Limited by cable length; ideal for localized areas \cite{b10, b12, b14} & Wide range, suitable for large geographic areas \cite{b11, b12, b13, b14, b15}\\
\hline
Application Suitability & Ideal for mission-critical operations (SCADA, substation automation) \cite{b10, b12}& Suited for flexible and remote applications (e.g., Remote metering) \cite{b14, b15}\\
\hline
Preferred For & Critical, real-time tasks requiring high reliability and low latency \cite{b10, b12}& Applications needing rapid deployment and scalability \cite{b14, b15}\\
\hline
\end{tabular}
\end{table}

\section{Emerging Use-cases and Service Requirements}
\label{sec:useserv}

\subsection{Emerging Use Cases}
Emerging use cases in SG are responding to the increasing complexity of emerging energy systems, a shift influenced by the integration of renewable energy sources. The growth of decentralized energy generation, and the rising prominence of prosumers. These use cases utilize advanced communications and information technologies, AI, and real-time data analytics to address the challenges encountered by modern energy grids or SG. We conducted a comprehensive research study and identified and classified some list of example SG use cases into several categories, as shown in Table \ref{tab:sgusecase}.

We describe the following SG use cases:
\begin{itemize}
\item{Primary Frequency Control}
\item{Smart Distributed Voltage Control}
\item{Smart Distributed Feeder Automation}
\item{Smart and Intelligent and Autonomous Monitoring}
\item{High-Speed Current Differential Protection}
\item{Smart and Intelligent Grid Millisecond-Level Precise Load Control}
\item{Distributed Energy Storage System}
\item{Smart and Intelligent Metering}
\item{Smart Distributed Transformer Terminal}
\item{DERs and Microgrid}
\item{Real-Time Fault Detection and Self-Healing}
\item{Predictive Maintenance}
\item{DERs Coordination}
\item{Identification of Fault Location}
\item{Differentiated QoS for Secured Smart Energy Data}
\item{Proactive Climate Resilience}
\item{Emergency Response and Disaster Recovery System}
\item{Automated Load Shedding}
\item{Edge Cloud Driven Data Acquisition}
\item{Intelligent and Autonomous Load Balancing}
\item{Advance Demand and Response}
\item{AI/ML-Based Distribution Grid Load and Generation Prediction}
\item{AI/ML-based Falling Conductor Detection with Root
Cause and Action}
\end{itemize}

\begin{table}[!t]
\caption{Categories of SG Use Cases \cite{b4, b6, b20, b21, b22, b24, b34, b35}\label{tab:sgusecase}}
\centering
\begin{tabular}{|p{45pt}|p{75pt}|p{90pt}|}
\hline
\textbf{Category} & \textbf{Description} & \textbf{Use-Cases} \\
\hline
Reliability, Resilience, and Automation & Ensures uninterrupted operations, fast fault recovery, and resilience to cyber or environmental disruptions. & Real-time fault detection, fault location identification, AI-based fault prediction, smart feeder automation, differential protection, transformer monitoring, UAV-based monitoring, millisecond-level load control, automated load shedding, climate resilience, and disaster recovery \cite{b4, b20, b21, b22, b24} \\
\hline
DER Integration and Optimization & Integrates renewable and distributed energy systems for efficient, optimized power flow. & Primary Frequency control, smart voltage control, energy storage management, renewable forecasting and optimization, DER coordination, microgrid control, VPP operations \cite{b4, b20, b21, b22, b24} \\
\hline
Consumer Engagement and Demand Management & Enables consumers to manage usage, participate in markets, and support grid balance. & Advanced demand response, demand-side optimization, smart metering, peer-to-peer trading, home energy management, carbon monitoring and reporting \cite{b4, b20, b21, b24} \\
\hline
Grid Stability and Control & Maintains voltage and frequency stability during dynamic grid conditions. & AI-based stability analysis, autonomous load balancing, smart voltage/frequency regulation \cite{b4, b20, b24} \\
\hline
Cybersecurity and Data Integrity & Protects grid data and infrastructure from cyber threats. & Cybersecurity intelligence, emergency response, disaster recovery systems \cite{b4, b6, b20, b24} \\
\hline
Data Processing and Edge Computing & Enhances local decision-making via real-time data analysis at the grid edge. & AI/ML for load prediction, predictive maintenance, falling conductor detection, digital twin modeling, edge PMUs \cite{b24, b34, b35} \\
\hline
EV Integration and Management & Manages charging demand and ensures grid reliability during EV adoption. & EV charging control, grid congestion management \cite{b4, b6, b20} \\
\hline
\end{tabular}
\end{table}

Each use case is summarized below with an explanation of its significance for SG.

\begin{itemize}
\item{\bf Primary Frequency Control:} 
Primary Frequency Control stabilizes grid frequency by dynamically balancing supply and demand through automatic generator adjustments. It prevents cascading failures and maintains system synchronization. URLLC plays a critical role in real-time data sharing for accurate frequency stabilization \cite{b20}. Integration with DERs and energy storage improves operational efficiency, especially in grids dominated by renewables. Advanced automation systems with predictive analytics ensure quick response times, compensating for frequency deviations caused by variable energy sources. This capability is essential for maintaining grid reliability, allowing grids to support a higher share of renewables \cite{b20, b22}. 

\item{\bf Smart Distributed Voltage Control: }
Smart Distributed Voltage Control ensures optimal voltage levels across power systems by dynamically regulating reactive power. It uses advanced devices like smart inverters and capacitor banks to stabilize voltage in grids with variable renewable energy sources. IoT-enabled sensors and edge computing facilitate real-time monitoring and decentralized control of voltage fluctuations. This will help to reduce energy losses, protect grid assets, and ensure consistent power quality for consumers. This use case is essential for both rural and urban grids, enhancing the operational efficiency of decentralized energy systems \cite{b20, b22}.

\item{\bf Smart Distributed Feeder Automation:}
Smart Distributed Feeder Automation enhances the reliability of distribution networks by streamlining processes such as fault detection, isolation, and restoration. This not only minimizes the duration of outages but also improves load balancing. Advanced sensors and real-time communication facilitate dynamic feeder reconfiguration \cite{b22}. By integrating AI-driven analytics and predictive maintenance models, feeder automation optimizes system performance and supports the seamless integration of DERs. The capability of this application is crucial for modern SG to achieve resilience and operational efficiency \cite{b20, b22, b23, b24}. 

\item{\bf Smart, Intelligent, and Autonomous Monitoring:}
Smart, Intelligent, and Autonomous Monitoring refers to real-time monitoring, security, and operational management of SG in generation, transmission, and distribution. This utilizes the use of advanced technologies such as high-resolution cameras, IoT-enabled sensors, environmental sensors, autonomous UAVs, AI for early anomaly detection and predictive maintenance, and secure communication networks to improve the reliability, efficiency, and resilience of SG. It ensures the safety and reliability of SG infrastructure by preventing unauthorized access, detecting faults, and allowing predictive maintenance \cite{b24}.

\item{\bf High-Speed Current Differential Protection:}
High-speed current differential protection isolates transmission line fault within milliseconds by comparing current measurements at both ends of a line. Synchronized relays and high-speed communication networks ensure precise and rapid fault identification \cite{b20}. It prevents cascading outages and equipment damage, enhancing grid reliability in high-voltage environments. Advanced relays and 5G cellular communication can make this protection mechanism indispensable for a SG \cite{b20}.

\item{\bf Smart and Intelligent Grid Millisecond-Level Precise Load Control:}
Smart and Intelligent Grid Millisecond-Level Precise Load Control serves as the fundamental use case within the SG system. In the event of a significant fault in HVDC (High-Voltage Direct Current) transmission. This is employed to quickly eliminate non-essential, interruptible loads, like EVs charging stations and non-continuous power supplied in manufacturing facilities and dynamically adjusts energy consumption to balance supply-demand fluctuations in renewable-heavy grids. It ensures grid stability by instantly balancing supply and demand, mitigating power overflows, and preventing system-wide outages \cite{b20, b24}. This use case can be essential for integrating renewable energy sources and managing distributed energy systems.

\item{ \bf Distributed Energy Storage System:}
Distributed energy storage is a powerful solution that includes various forms such as solar energy, wind energy, fuel cells, and gas and a combination of multiple forms. The energy is generated, stored and supplied directly at the user’s site or in close proximity, ensuring efficient and reliable access to power. These systems generate and store extra energy for use during peak demand, outages, or emergencies, improving grid resilience and supporting renewable energy integration. In case of severe weather that disrupts the larger power grid, It can operate independently to create islands or microgrids, providing emergency power to critical facilities like hospitals and transportation hubs. Storage systems in regulating frequency, balancing load and supplying backup power \cite{b24}. Integrating these systems with advanced communication networks such as 5G networks optimizes their performance in a decentralized grid \cite{b24}.

\item{ \bf Smart and Intelligent Metering:}
Smart and Intelligent Metering utilizes a Smart Metering Infrastructure system, to allow two-way communication between utilities and consumers, facilitating real-time monitoring of energy usage. It improves demand response, enhances load forecasting, and supports dynamic pricing. The Smart and Intelligent Metering Infrastructure promotes energy efficiency and permits consumer engagement in the energy market \cite{b4, b20, b24}.

\item{ \bf Smart Distributed Transformer Terminal:}
Smart Distributed Transformer Terminals allow real-time monitoring, control, and management of transformers, allowing for predictive maintenance, automated fault detection (e.g., falling conductor detection), and optimization of distributed load across transformers to increase the efficiency of energy distribution \cite{b20, b24}. According to \cite{b6}, it is essential to implement secure communication protocols to safeguard data integrity and enhance the reliability of transformer operations. These terminals contribute to improved grid stability and reduce operational costs.

\item{\bf DERs and Microgrid:}
DERs refer to a small-scale generation or storage units of electricity, such as rooftop solar panels, wind turbines, and battery systems, that are connected to the distribution grid. Microgrid integrate these DERs, inspiring them to operate as localized grids that can function independently (i.e., in an “islanded mode” or in connection with the main grid). Microgrids enhance both resilience and flexibility. Microgrids significantly improve energy reliability during outages or emergencies and facilitate the deployment of renewable energy sources \cite{b20, b22}.

\item{\bf Real-Time Fault Detection and Self-Healing:}
The increasing complexity of modern power systems has necessitated the adoption of advanced fault management solutions. This is an advanced operational feature that uses continuous monitoring, rapid fault analysis, and autonomous restoration processes to ensure grid reliability and resilience. By integrating IoT sensors, smart meters, and real-time communication technologies, the system can detect anomalies and faults as they occur. AI/ML algorithms analyze performance data to diagnose the location and nature of faults, allowing swift isolation through automated sectionalizing switches or similar mechanisms. Following fault isolation, self-healing capabilities reroute power supply dynamically, often utilizing DERs and microgrids, to maintain service continuity. This mechanism reduces outage duration significantly and prevents cascading failures across the grid \cite{b20, b21}. The key features include low-latency response times, typically in milliseconds to seconds, ensuring immediate mitigation actions. This use case enhances grid reliability and operational efficiency by minimizing manual intervention and maintenance costs. It is critical for urban power grids, critical infrastructure such as hospitals, and systems integrating renewable energy, where rapid fault recovery is essential to maintaining stability and service \cite{b22, b23}. It serves as a backbone of resilient SG systems, aligning with modern demands for adaptive, autonomous, and efficient energy networks \cite{b24}.

\item{\bf Predictive Maintenance:}
Predictive Maintenance is transforming how utilities manage SG infrastructure by shifting from reactive to proactive interventions. Through continuous real-time monitoring and analysis of IoT sensor data, advanced machine learning algorithms detect patterns and anomalies that signal potential failures before they occur. This can eliminate the inefficiencies of traditional maintenance schedules, focusing resources precisely on where they are needed.

A transformative element of this use case is the digital twin, a virtual model of physical grid elements. These replicas simulate real-time operations, allowing utilities to predict equipment behavior under different conditions. For example, transformers and substations, once reliant on costly inspections, are now monitored continuously, with alerts triggered only when performance deviates from optimal levels \cite{b20, b23}. The advantages are clear: reduced downtime, extended equipment lifespan, and significant cost savings. GE Vernova’s SmartSignal platform is a leading example of this. Using over 350 different digital twin blueprints to provide predictive insights across turbines, pumps, and entire plants, saving over \$1.6 billion for clients \cite{b35}. These digital twins ingest diverse operational data from Human-Machine Interface (HMI)/SCADA systems to ambient temperature, enabling precise real-time diagnostics. Notably, their Original Equipment Manufacturer (OEM)-agnostic nature ensures broad applicability across hybrid industrial fleets \cite{b35}.

This also enhances grid reliability, ensuring uninterrupted energy delivery and fostering confidence in integrating renewable energy sources. In critical infrastructures such as hospitals or data centers, it ensures consistent power supply, minimizing disruptions \cite{b21, b24}. Moreover, predictive maintenance is not just about efficiency, it is about resilience. By identifying vulnerabilities and addressing them preemptively, utilities build smarter, more adaptive grids capable of meeting modern challenges, including rising energy demand and climate uncertainties. This marks a shift from problem-solving to innovation, where the grid of tomorrow is not only reliable but anticipates and adapts to future needs \cite{b22}.

\item{\bf{DERs Coordination:}}
DERs Coordination involves integrating and managing various DERs, such as solar panels, wind turbines, battery storage, and microgrids, with the power system. It ensures that DERs operate harmoniously with the grid, contributing to reliable energy supply, efficient energy distribution, and grid stability. DER coordination requires advanced information and communication systems, real-time data sharing, and intelligent control algorithms to optimize energy production, balance supply and demand, and prevent grid disturbances. DERs are often geographically dispersed and vary in size, generation capacity, and operational characteristics. Effective coordination includes monitoring DER performance, forecasting generation and consumption, and implementing dynamic control strategies. This process allows for efficient energy distribution and enhanced grid resilience \cite{b20}.

\item{\bf Identification of Fault Location:}
Identification of Fault Location is a method of detecting and pointing out the exact locations of faults within a power grid to ensure quick and efficient resolution of issues. Faults such as short circuits, line failures, or equipment malfunctions can disrupt the normal flow of electricity, leading to power outages, system instability, or equipment damage. This feature employs advanced sensors, a PMU, and automated systems to continuously monitor the grid for any abnormalities. Upon detecting a fault, real-time data regarding current and voltage fluctuations is transmitted to a control system, where advanced algorithms analyze the information to accurately identify the fault’s location. By allowing quick isolation and repair faults, this approach reduces downtime, enhances grid reliability, and lowers operational costs. Identification of the fault location is quite critical in modern SG, which involve complex topologies and DERs, as it ensures efficient fault resolution and supports the overall resilience of the SG \cite{b20}.

\item{\bf Differentiated QoS for Secured Smart Energy Data:}
Differentiated QoS for Secure Energy Data ensures that encrypted data communication within the SG system is prioritized based on the criticality of the SG applications or use cases. Differentiated QoS allocates specific service levels to the data streams, such as fault detection signals, distributed automation, distributed generation, metering data, and energy demand-response communications, ensuring timely and secure delivery without compromising encryption. In SG, encryption safeguards data integrity and confidentiality, but it can obscure traffic characteristics that traditional QoS mechanisms rely on. Differentiated QoS for encrypted data influences advanced network policies and traffic classification techniques to maintain efficiency while upholding security standards. This differentiation is important to maintain grid stability, optimize resource allocation, and enhance operational security in increasingly dynamic and data-intensive energy systems \cite{b24}.

\item{\bf Proactive Climate Resilience:}
The Proactive Climate Resilience use case in SG focuses on preparing for and mitigating the impacts of climate-induced events. Through advanced monitoring, predictive analytics, and adaptive responses, by utilizing IoT sensors, 5G communication, and AI-based analytics to monitor environmental conditions in real-time in generation, transmission, distribution, and prosumers. To forecast extreme weather event, and predict potential disruptions to SG infrastructure, allowing preemptive responses, ensuring reliable energy delivery and sustainability \cite{b24}. 

\item{\bf Emergency Response and Disaster Recovery System:}
The Emergency Response and Disaster Recovery System can be an essential SG application that ensures quick response, relief, and recovery during natural disasters such as hurricanes, floods, earthquakes, heavy snowfall, wildfires, cyberattacks, or other emergency situations. By leveraging real-time information, automated diagnostics, and advanced communication technologies. This system will help utilities assess damage, isolate faults, and restore power more efficiently. It also supports coordination with emergency service and critical infrastructure to prioritize recovery operations and ensure public safety \cite{b4, b20, b24}.

\item{\bf Automated Load Shedding:}
Autonomous Load Shedding facilitates the automatic and intelligent disconnect of non-critical electric loads such as air conditioning, washing machines, and non-priority electric vehicle charging stations in response to grid disturbances, such as sudden overloading or generation failures. It functions independently, employing real-time monitoring, AI-driven decision-making algorithms and ultra-low latency communication systems. It is designed to maintain grid stability and mitigate the risk of cascading failures \cite{b20, b24}.

\item{\bf Edge Cloud-Driven Data Acquisition:}
The Edge Cloud Driven Data Acquisition use case revolutionizes SG monitoring by utilizing 5G edge cloud capabilities for efficient and cost-effective phasor measurement. Traditionally, PMUs are expensive and difficult to scale. The edgePMU addresses this limitation by replacing traditional PMUs with lower-cost field devices that collect key grid parameters such as voltage, current, frequency, and the rate of change of frequency. These measurements are transmitted via communication networks such as 5G networks to the edge cloud, where advanced processing occurs in real-time. The outputs, including synchronized phasor data, are essential for services like fault detection, and other grid management tasks. This ensures high scalability and flexibility, as the processing and service upgrades occur dynamically in the edge cloud without relying on specialized and expensive on-site hardware. The edgePMU is particularly optimized for stationary field devices operating within the power distribution grid. By integrating 5G and edge computing, edgePMU enhances real-time data acquisition and grid stability, paving the way for an intelligent and resilient SG infrastructure \cite{b24}.

\item{\bf Intelligent and Autonomous Load Balancing:}
Intelligent and Autonomous Load Balancing in the SG provides real-time automatic and intelligent load distribution or required adjustment across the SG network to maintain grid stability, optimize resource utilization, and manage fluctuating demand and supply situations. It prevents infrastructure overloading, enhances energy distribution, and supports seamless integration of renewable energy sources such as solar energy and wind energy. This use case ensures that SG infrastructure remains operational, overloading is avoided, and efficiency of grid is maximized \cite{b20, b24}.

\item{\bf Advanced Demand and Response:}
Advanced Demand and Response serves as a dynamic energy management system that uses real-time communication, automation, and pricing information to manage electricity consumption based on grid conditions. It enables utilities and grid operators to effectively balance supply and demand by encouraging consumers, whether residential, commercial, or industrial, to adjust their energy usage during peak demand times or in response to grid emergencies. It is essential for enhancing grid flexibility, supporting the integration of renewable energy by aligning energy consumption with generation patterns such as utilizing solar power during daylight hours, empowering consumers to engage actively in energy markets, reducing dependence on fossil fuel-powered peaking power plants, and contributing to a more sustainable, resilient and efficient energy system \cite{b4, b20, b24}.

\item{\bf AI/ML-Based Distribution Grid Load and Generation Prediction:}
The AI/ML-based Distribution Grid Load and Generation Prediction use case forecasts future electricity consumption and generation to provide the Distributed System Operator (DSO), insights into grid operating conditions in various time horizons. Historical data of electricity consumption and generation at customer, generator, or substation levels can be utilized. Additionally, external factors such as weather conditions and temperature can be important inputs. Using advanced AI/ML algorithms, the possible impacts of these variables on electricity performance can be accurately assessed. This system can predict the electricity conditions a day in advance with time resolutions, e.g. 15 minutes, 30 minutes, hourly, and daily. While the short-term forecast (e.g., an hour in advance) is more reliable due to the real-time grid status updates, long-term forecasts support strategic planning and grid reinforcement decisions by covering the hourly, seasonal, monthly, or yearly patterns. This use case enhances the understanding of future grid conditions for DSO, provides preventive measures to improve reliability, and completes long-term planning for effective grid infrastructure management \cite{b24}.

\item{\bf AI/ML-based Falling Conductor Detection with Root Cause and Action Recommendation:}
AI/ML-based Falling Conductor Detection with Root Cause and Action Recommendation use case consists of three major functions: Falling Conductor Detection, Root Cause Analysis, and Action Recommendation. It ensures real-time detection of falling conductors and de-energizes them before they reach the ground and automatically isolates the affected line, determines the actual reason for the cause, and gives the possible action recommendation to fix it along with the estimated healing time. This use case collects and merges diverse real-time data streams, including PMU statistics such as voltage, current, temperature, and the rate of change in electrical impedance. It also incorporates data from environmental sensors to monitor weather conditions such as wind speed, temperature, and humidity. In addition, cameras and UAV-mounted cameras in remote locations capture images and videos of the surrounding environment. These multi-source data streams are analyzed using advanced AI/ML algorithms to support situational awareness and predictive decision-making. This use case enhances public safety, reduces the risk of wildfires, and ensures grid reliability by minimizing infrastructure damage and outage durations, leading to significant cost savings \cite{b34}. 

\item{\bf Distributed Energy Storage Management:}
Distributed Energy Storage Management (DESM) can be the centralized control system for managing distributed energy storage devices (DEDs) within decentralized power generation networks. These systems, including batteries, fuel cells, and other storage technologies, are crucial to stabilizing SG as energy generation shifts towards localized and renewable sources. In distributed grids, prosumers lead to bidirectional energy flows. These dynamic flows introduce complexities such as varying current direction at different grid locations and times, necessitating sophisticated management systems \cite{b20}.
Information exchange is a key aspect of DESM. Communication between DEDs and the Distributed Energy Storage Management Platform (DESMP) involves real-time data transfer. DEDs collect and periodically transmit critical energy metrics, including battery status, charge/discharge levels, and energy alarms. Using high-speed 5G networks, this data can be sent to the DESMP, which monitors the working status of DEDs, configures their energy parameters, and controls their operating modes. The DESMP also issues control commands such as adjust load characteristics, to ensure a flexible and stable electricity grid \cite{b21}. One of the defining features of DEDs is their plug-and-play design, simplifying integration into existing energy systems. This capability combined with the DESMP’s real-time coordination, allows for seamless operations in decentralized environments. By optimizing energy storage and retrieval, DESM enhances grid flexibility and operational efficiency. DESM addresses the complexities introduced by renewable energy variability and decentralization, making it a cornerstone of adaptive, reliable, and efficient SG systems \cite{b23}.
\end{itemize}

\subsection{Service Requirements for Selected Use Cases}
We have carried out a comprehensive study and identified the service requirements in Table \ref{tab:sgucservicerequirements} based on \cite{b20, b21, b22,b24} to ensure the effective operation of the defined SG use cases. Each use case may apply to one or more than one domain. The service requirements include various parameters such as availability, latency, data rate, and Transmission Time Interval (TTI), many of which are quite stringent. Consequently, it is essential to establish an integrated communication network that meets these requirements to effectively implement these use cases.

\begin{table*}[htbp]
\caption{SG Use Cases Service Requirements\label{tab:sgucservicerequirements} \cite{b20, b21, b22,b24}}
\centering
\resizebox{\textwidth}{!}{%
\begin{tabular}{|p{3cm}|p{1.5cm}|p{0.9cm}|p{1.2cm}|p{1.2cm}|p{1.5cm}|p{1.5cm}|p{1.0cm}|p{1.0cm}|p{1.5cm}|p{0.8cm}|p{0.8cm}|p{0.8cm}|}
\hline
\bfseries Use Case &
\rotatebox{90}{\bfseries Domain Level} &
\rotatebox{90}{\bfseries Voltage Level} &
\rotatebox{90}{\bfseries Availability (\%)} &
\rotatebox{90}{\bfseries Reliability (\%)} &
\rotatebox{90}{\bfseries Latency} &
\rotatebox{90}{\bfseries Data Rate} &
\rotatebox{90}{\bfseries Message Size} &
\rotatebox{90}{\bfseries TTI} &
\rotatebox{90}{\bfseries Connection Density} &
\rotatebox{90}{\bfseries Service Area (km\textsuperscript{2})} &
\rotatebox{90}{\bfseries Time Sync Accuracy} &
\rotatebox{90}{\bfseries Jitter} \\
\hline
Primary Frequency Control & Gen, Trans, Dist, Cons & H, M, L & 99.9999 \cite{b20}& TBD \cite{b20} & $\sim$50 ms \cite{b20, b22} & - & $\sim$100 byte \cite{b20, b22} & $\sim$50 ms \cite{b20}\cite{b22} & $\leq$ 100000 \cite{b20}\cite{b22} & $\leq$ 100000 \cite{b20}\cite{b22} & - & - \\
\hline
Smart Distributed Voltage Control & Gen, Dist, Cons  & H, M, L & 99.999 \cite{b20} & TBD \cite{b20} & $\sim$100 ms \cite{b20}\cite{b22} & - & $\sim$100 byte \cite{b20, b22} & $\sim$200 ms \cite{b20}\cite{b22} & $\leq$ 100000 \cite{b20, b22} & $\leq$ 100000 \cite{b20}\cite{b22} & - & - \\
\hline
Smart Distributed Feeder Automation & Dist & M, L & 99.999 \cite{b20}\cite{b22} & 99.999 \cite{b24} & 2 ms \cite{b20} & 2 Mbps to 10 Mbps \cite{b20} & - & 1s (Normal), 2ms (Fault) \cite{b20} & 54/km\textsuperscript{2}, 78/km\textsuperscript{2} \cite{b20} & Several \cite{b20} & 10$\mu$s \cite{b24} & 50$\mu$s \cite{b24} \\
\hline
Smart and Intelligent Substation Monitoring & Trans, Dist & M, L & - & - & - & upto 5 Mbps \cite{b24} & - & - & upto 100 \cite{b20}\cite{b24} & $\leq$ 20/km\textsuperscript{2} \cite{b20}\cite{b24} & $\leq$10-20$\mu$s \cite{b20}\cite{b24} & - \\
\hline
High-Speed Current Differential Protection & Trans, Dist & H, M, L & $>$99.999 \cite{b20, b21, b24} & 99.9999 \cite{b20, b21, b24} & 5 ms \cite{b20}\cite{b24}& 2.5 Mbps\cite{b20, b21, b24} & $<$245 byte \cite{b20, b21, b24}  & 1 ms \cite{b20}\cite{b24} & $\leq$ 100/km\textsuperscript{2} \cite{b20, b21} & - & 10$\mu$s \cite{b21, b22, b24}& $<$160$\mu$s \cite{b21} \\
\hline
Smart and Intelligent Grid Millisecond-Level Load Control & Trans, Dist, Cons & H, M, L & 99.9999 \cite{b20}\cite{b22}& - & $<$50 ms \cite{b20}\cite{b22}& 0.59 Kbps, 28 Kbps \cite{b20} & $<$100 byte \cite{b20} & - & 10–100/km\textsuperscript{2} \cite{b20}& - & - & - \\
\hline
Distributed Energy Storage System & Gen, Dist, Cons & M, L & $>$99.99 \cite{b20}\cite{b24} & - & 10 ms \cite{b20}\cite{b24} & UL $>$ 640 Mbps, DL $>$ 5 Gbps \cite{b20}\cite{b24} & - & UL $<$ 10 ms \cite{b20}\cite{b24} & $>$100/km\textsuperscript{2} \cite{b20}\cite{b24} & Several \cite{b20}& - & - \\
\hline
Smart and Intelligent Metering & Gen, Dist, Con & M, L & $>$99.99 \cite{b20}\cite{b24} & - & $<$100 ms \cite{b20}\cite{b24} & UL $>$ 2 Mbps, DL $>$ 1 Gbps \cite{b20}\cite{b24} & - & - & $<$10000/km\textsuperscript{2} \cite{b20}\cite{b24} & - & - & - \\
\hline
Smart Distributed Transformer Terminal & Dist & M, L & $>$99.99 \cite{b20}\cite{b24} & - & 10 ms \cite{b20}\cite{b24} & $>$2 Mbps \cite{b20}\cite{b24} & - & - & 500 in service area \cite{b20}\cite{b24} & 100–500 m \cite{b20}\cite{b24} & - & - \\
\hline
DERs and Micro-Grids & Gen, Dist, Con & M, L & 99.9999 \cite{b20}\cite{b24} & - & $<$3 ms \cite{b20}\cite{b24} & 5.4 Mbps \cite{b20}\cite{b24} & 160 byte \cite{b20}\cite{b24} & $\leq$ 1 ms \cite{b20}\cite{b24} & - & - & - & - \\
\hline
Real-Time Fault Detection \& Self Healing & Trans, Dis, Cons & H, M, L & 99.9999 \cite{b20}\cite{b21} \cite{b22}& 99.999 \cite{b21} & $<$5 ms \cite{b20}\cite{b22} & 1.5 Mbps \cite{b20}\cite{b21} \cite{b22} & 1500 byte \cite{b20}& $\geq$ 1 ms \cite{b20} & 20 \cite{b20} \cite{b22}& 30 km $\times$ 20 km \cite{b20}& - & - \\
\hline
Identification of Fault Location & Trans, Dist & H, M, L & 99.9999 \cite{b20}\cite{b21} & 99.9999 \cite{b21}& 140 ms \cite{b21} & 100 Mbps \cite{b21} & - & - & 10/km\textsuperscript{2} \cite{b21}& - & 5$\mu$s \cite{b21} & 2 ms \cite{b21} \\
\hline
Predictive Maintenance & Gen, Trans, Dist, Cons & H, M, L & 99.99 \cite{b20} & TBD \cite{b20} & < Transfer interval value \cite{b20} & $\geq$ 1 Mbps \cite{b20} & upto 255 byte \cite{b20} & Several seconds \cite{b20} & $\leq$ 100000 \cite{b20} & - & - & - \\
\hline
DERs Coordination & Gen, Dist, Cons & M, L & 99.9999 \cite{b20} & - & $<$5 ms \cite{b20} & 1.5 Mbps \cite{b20} & 1500 byte \cite{b20} & $\geq$ 1 ms \cite{b20} & 20 \cite{b20} & 30 km $\times$ 20 km \cite{b20} & - & - \\
\hline
Differentiated QoS for Energy Data & Gen, Trans, Dist, Cons & M, L & 99.99 \cite{b24} & - & 2 s \cite{b24} & 5 Mbps \cite{b24} & - & - & - & - & - & - \\
\hline
Proactive Climate Resilience & Gen, Trans, Dist, Cons & H, M, L & - & - & - & - & - & - & - & - & - & - \\
\hline
Emergency Response and Disaster Recovery System & Trans, Dist, Cons & M, L & - & - & - & - & - & - & - & - & - & - \\
\hline
Automated Load Shedding & Dist & M, L & - & - & - & - & - & - & - & - & - & - \\
\hline
Intelligent and Autonomous Load Balancing & Dist & M, L & - & - & - & - & - & - & - & - & - & - \\
\hline
Edge-Driven Data Acquisition & Dist & M, L & - & $>$1 yr \cite{b24} & $<$33 ms \cite{b24} & 20 Mbps \cite{b24} & - & $\leq$ 1 ms \cite{b24} & - & - & - & - \\
\hline
Advance Demand Response & Dist & M, L & 99.99 \cite{b24} & - & 500 ms \cite{b24} & 100 Kbps \cite{b24} & - & - & - & - & - & - \\
\hline
AI/ML-based Distribution Grid Load and Generation Prediction & Gen, Dist, Cons & M, L & 99.9 \cite{b24} & - & Not Critical \cite{b24} & 100 bps \cite{b24} & $<$1000 byte \cite{b24} & 1 min \cite{b24} & $<$30/km\textsuperscript{2} \cite{b24} & $\leq$ 100000 km\textsuperscript{2} \cite{b24} & - & - \\
\hline
AI/ML-based Falling Conductor Detection & Trans, Dist & M, L & 99.9999 \cite{b20} \cite{b24} & - & $<$10 ms \cite{b22} \cite{b24}& - & - & - & 100/km\textsuperscript{2} \cite{b22}& - & 1$\mu$s \cite{b22}  & - \\
\hline 
\multicolumn{13}{l}{\footnotesize \textit{Note:} Gen: Generation, Trans: Transmission, Dist: Distribution, Cons: Consumer, H: High, M: Medium, L: Low, UL: Uplink; DL: Downlink; TBD: To Be Determined; --: Not Specified.} \\
\end{tabular}}
\end{table*}

\begin{itemize}
\item{Availability: }
It refers to the proportion of time a network or service is operational and accessible.
\item{Latency:}
The maximum time delay between sending a signal and receiving the response. 
\item{Data Rate:}
The speed of data transmission over a communication network.
\item{Message Size:}
The amount of data contained in a single communication exchange or message.
\item{TTI:}
TTI is the minimum time between the transmission of two consecutive data packets.
\item{Connection Density:}
The maximum number of devices that can connect within a specific area.
\item{Service Area:}
The geographical region where a specific service is provided.
\item{Time Synchronous Accuracy:}
It refers to the precision with which systems align clocks or timestamps across devices.
\item{Jitter:}
It refers to the variation in the delay (latency) of transmitted data packets over a network.
\end{itemize}

Analyzing the data shown in Table \ref{tab:sgucservicerequirements} highlights notable trends and significant diversity in service requirements. Mission-critical domains, particularly in protection and control, demand exceptionally high levels of availability, often exceeding 99.999\%. There is an observable inverse relationship between latency requirements and data rates. Protection use cases such as Smart Distributed Feeder Automation and High-Speed Current Differential Protection necessitate extremely low latencies of up to 2 ms, accompanied by time synchronization within 10 $\mu$s, despite relatively modest data payloads. Conversely, the DESM stands out as a bandwidth-demanding system, requiring downlink rates exceeding 5 Gbps while still adhering to low-latency targets of around 10 ms. Connection density requirements also show considerable variation, soaring up to 100,000 devices per km\textsuperscript{2} for distributed voltage and frequency control. This heterogeneity indicates that a static network configuration is inadequate; the infrastructure must accommodate both massive machine-type communications and ultra-reliable low-latency services. Additionally, for several use cases, AI/ML will be highly valuable.

The requirements outlined exceed the capabilities of current 4G/LTE technology. High-Speed Current Differential Protection and millisecond-level load control demand latency of approximately a few milliseconds, an availability rate of \~99.999\%, and time-synchronized measurements (see Table \ref{tab:sgucservicerequirements}). In contrast, 4G LTE generally presents higher and variable latency (ranging from 20 to 100 ms), experiences diminished reliability during congestion, and lacks optimization for densely packed, time-critical uplinks from protection devices. In particular, the radio interface limitations due to the longer TTI resulting from narrower subcarrier spacing and lack of a comprehensive network slicing framework to meet QoS requirements restrict achievable latency in 4G LTE. This scenario underscores the need for 5G/6G URLLC and flexible uplink scheduling to support mission-critical applications within the SG.

\section{Capabilities of 5G and 6G for Emerging SG}
\label{sec:capab}
\begin{figure}[!t]
\centering
\includegraphics[width=3.4in]{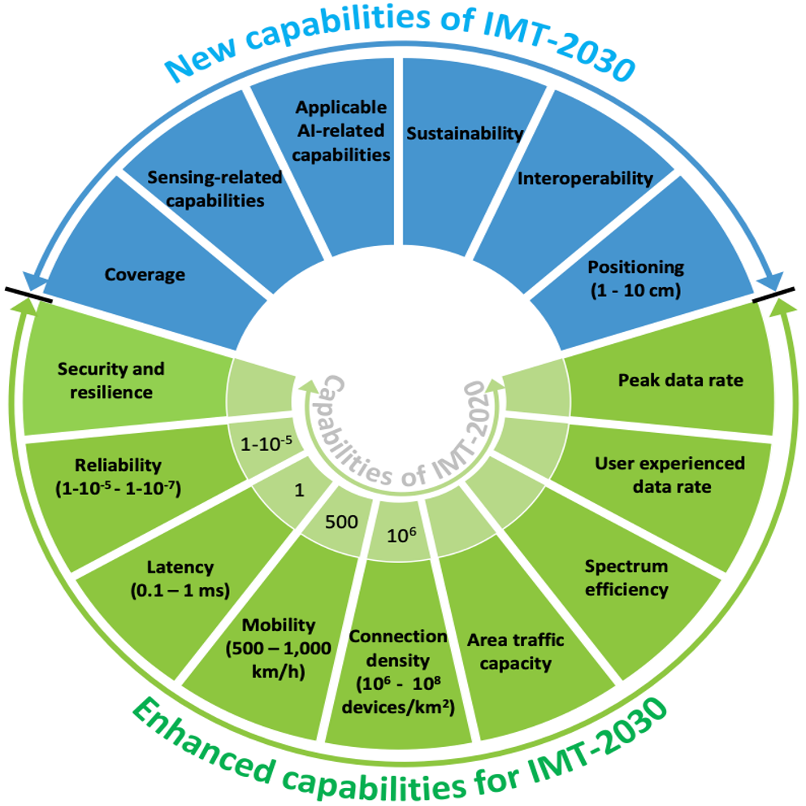}
\caption{Capabilities of 6G defined by ITU under IMT-2030 \cite{b23}}
\label{fig:ITU2030}
\end{figure}

In this section, we examine the capabilities of 5G and 6G in the context of the service requirements discussed in Section \ref{sec:useserv}. In Fig. \ref{fig:ITU2030}, the green section illustrates the enhanced capabilities of 6G (denoted as IMT-2030) relative to those of 5G (denoted as IMT-2020), as well as the new capabilities of 6G highlighted as blue. The enhanced capabilities are specified in the form of Peak Data Rate, User Data Rate, Spectral Efficiency, Area Traffic Capacity, Connection Density, Mobility, Latency, Reliability, Security, and Resilience. While new capabilities are coverage, sensing, AI capabilities, sustainability, interoperability, Positioning introduced in 6G. Related KPIs are defined below \cite{b23, b25}. 
\begin{itemize}
\item{Peak Data Rate:}
The highest possible data transmission rate achievable under flawless conditions. It indicates maximum network throughput.

\item{User Data Rate:}
The achievable data rate available to users across the network’s coverage area. It is essential for ensuring consistent user experience.

\item{Spectrum Efficiency:}
The amount of data transmitted per unit of spectrum resource and per cell. This defines how effectively the spectrum is utilized.

\item{Area Traffic Capacity:} The total data throughput served within a specific geographic area. It is shows the network density capabilities.

\item{Connection Density:} The total number of connected devices per square kilometer. It is a measure of the network's ability to support IoT and dense device environments.

\item{Mobility:} The maximum speed at which reliable communication is maintained. It is essential for supporting fast-moving devices.

\item{Latency:} The time delay for data to travel from source to destination. It is critical for real-time applications such as URLLC.

\item{Reliability:} The probability of successful data transmission within a defined time frame. It is important for mission critical services.

\item{Security and Resilience:} The ability to protect against cyber threats and maintain reliable operation under disruption, including natural or man-made events.

\item{Coverage:} The ability to provide users with access to communication services within a specific area. This is defined as the cell edge distance of a single cell through link budget analysis.

\item{Sensing:} The ability to provide functionalities such as range, velocity, and angle estimation, object detection, localization, and imaging and mapping using the radio interface. These capabilities can be evaluated based on accuracy, resolution, detection rate, and false alarm rate.

\item{AI Capabilities:} The integration of AI into the network for optimization, autonomous management, and enhanced decision making.

\item{Sustainability:} Efforts to minimize environmental impacts by improving energy efficiency and using sustainable technologies across the network infrastructure.

\item{Interoperability:} The ability of the network to seamlessly integrate with other systems, including non-terrestrial networks and legacy technologies.

\item{Positioning:} The ability to accurately determine the location of a device with centimeter-level precision, essential for navigation and IoT applications.
\end{itemize}

\begin{table*}[!t]
\caption{Capabilities of 5G and 6G for SG Use-Cases \cite{b23, b25, b26, b36}}
\label{tab:5g6g_capabilities}
\begin{tabular}{|p{75pt}|p{180pt}|p{50pt}|p{90pt}|p{50pt}|}
\hline
\bf{KPIs} & \bf{Description} & \bf{5G} & \bf{6G} & \bf{SG Requirements} \\
\hline
Peak Data Rate & Maximum achievable data rate under ideal conditions & 20 Gbps \cite{b25, b36} & up to 200 Gbps \cite{b25, b23} & - \\
\hline
User Data Rate & Achievable data rate ubiquitously available to users in the coverage area & 100 Mbps \cite{b25, b36} & up to 500 Mbps \cite{b25, b23} & 5 Gbps (from Table \ref{tab:sgucservicerequirements})\\
\hline
Spectrum Efficiency & Data throughput per unit spectrum resource per cell & 3× over 4G \cite{b25, b36} & Up to 3× over 5G \cite{b25, b23} & - \\
\hline
Area Traffic Capacity & Total traffic served per geographic area & 10 Mbps/m\textsuperscript{2} \cite{b25, b36} & Up to 50 Mbps/m\textsuperscript{2} \cite{b25, b23} & - \\
\hline
Connection Density & Total number of connected devices per square kilometer & 1 million \cite{b25, b36} & up to 100 million \cite{b25, b23} & 0.1 million (from Table \ref{tab:sgucservicerequirements})\\
\hline
Mobility & Maximum speed supported while maintaining reliable communication & Up to 500 km/h \cite{b25, b36} & Up to 1000 km/h \cite{b25, b23} & Stationary \\
\hline
Latency & Time taken for a packet to travel from source to destination over the air interface & 1 ms \cite{b25, b36} & 0.1 ms \cite{b25, b23} & 2 ms (from Table \ref{tab:sgucservicerequirements}) \\
\hline
Reliability & Probability of successful transmission within a predefined time duration & 99.999\% \cite{b25, b36} & 99.9999\% \cite{b25, b23} & 99.999\% (from Table \ref{tab:sgucservicerequirements})\\
\hline
Security and Resilience & Ability to protect data and continue operation under disruptions & Improved vs 4G \cite{b36} & Quantum-based, AI-driven security \cite{b23} & 100\% \cite{b6}\\
\hline
Coverage & Ability to provide communication services across different areas & Urban and Rural \cite{b36}  & Global, including non-terrestrial \cite{b25, b23} & - \\
\hline
Sensing & Ability to sense objects, detect motion, and provide situational awareness using radio signals & Limited \cite{b26, b36} & ISAC \cite{b25, b23} & - \\
\hline
AI Capabilities & Integration of AI for optimization and decision making & AI-augmented \cite{b26, b36} & AI-native \cite{b25, b23} & - \\
\hline
Sustainability & Energy efficiency and environmental impact of communication technologies & 10× more efficient than 4G \cite{b36} & 100× more efficient than 5G \cite{b25, b23} & - \\
\hline
Interoperability & Ability to work seamlessly across various network types and systems & Limited \cite{b36} & Extensive across terrestrial and non-terrestrial systems \cite{b25, b23} & - \\
\hline
Positioning & Ability to determine the position of connected devices with accuracy & Meter-level \cite{b25, b36} & Centimeter-level (1–10 cm) \cite{b23, b25} & - \\
\hline
\end{tabular}
\end{table*}

The 3GPP has completed Release 19 with 5G-Advanced enhancements in December 2025. The 3GPP is currently carrying out 6G studies in the ongoing Release 20, which will be finalized in June 2026. The 3GPP will begin its work on the initial specifications of 6G in Release 21 in the second half of 2027. Commercial products supporting a given 3GPP release typically become available in 18 to 24 months after the release completion. Hence, commercial 6G products that can meet most stringent SG use cases can be expected around 2030. The ITU has also aimed for 6G availability in the 2030 timeframe and has specific 5G requirements as IMT-2030 requirements.

Table \ref{tab:5g6g_capabilities} provides a comparison of the KPIs for 5G and 6G, emphasizing their alignment with the critical requirements for SG applications as outlined in Table \ref{tab:sgucservicerequirements}. These requirements encompass demanding criteria such as higher data rates, ultra-low latency, high availability, enhanced reliability, and the ability to support a large number of connections for various SG use cases. Both technologies support standard services and customized requirements combining these capabilities.

Although 5G URLLC supports latencies of approximately 1 ms, mission-critical applications, including High-Speed Current Differential Protection and Millisecond-Level Precise Load Control, exceed these limitations during periods of severe grid congestion. Ensuring time synchronization within 10 $\mu$s and achieving an availability of 99.9999\%, while simultaneously accommodating massive concurrent IoT traffic, necessitates the 0.1 ms latency and deterministic scheduling capabilities anticipated with 6G technologies.

\begin{table}[!t]
\caption{New Capabilities of 6G for SG Use-Cases \cite{b23, b25}\label{tab:newcab6g}}
\centering
\begin{tabular}{|p{60pt}|p{160pt}|}
\hline
\bf{New Capabilities } & \bf{SG Use Cases shown in Table \ref{tab:sgucservicerequirements}} \\
\hline
Coverage & 6G provides global coverage, connecting urban, rural, and remote areas \cite{b23, b25}; Ensures seamless communication for DERs, microgrids, and SG operations.; Supports real-time fault detection and energy management. \\
\hline
AI Capabilities & It enables real-time data analysis, autonomous decision-making, real-time optimization, continuous learning, and intelligent reconfiguration capabilities \cite{b23, b25}; Generative Al may play a crucial role in improving the SG use cases. It is building intelligent, adaptive, and resilient energy systems for the future. \\
\hline
Sensing & ISAC is a unique characteristic of 6G, highlighting its ability to merge communication and sensing into a single cohesive system, which supports advanced applications requiring real-time and precise information \cite{b23, b25}; ISAC can enhance fault detection, grid optimization, and DERs integration through its advanced sensing capabilities. \\
\hline
Sustainability & Integrate advanced energy-saving technologies, such as Near Zero Energy (NZE) communications, and optimize time, frequency, spatial, and power domains, along with renewable energy sources \cite{b23, b25}; Aim to reduce energy consumption and enhance the energy efficiency of communication and control devices within SG systems.; Utilize an open-source framework to promote and streamline the reuse of software and data related to SG technologies. \\
\hline
Interoperability & Seamless integration of heterogeneous networks (e.g., loT, Internet of Energy, Industrial loT).; Cross-system collaboration among devices and applications across different industries \cite{b23, b25}; Efficient deployment of multi-vendor solutions in a unified, interoperable ecosystem. \\
\hline
Positioning & 6G enables ultra-precise positioning for accurate fault detection and grid asset tracking \cite{b23, b25}.; High-precision Global Positioning System (GPS) and sensing improve grid maintenance and infrastructure inspections \cite{b23, b25}; Real-time positioning enhances outage management and resource optimization. \\
\hline
\end{tabular}
\end{table}
Several SG use cases will utilize AI/ML to analyze data in real time, helping the detection of patterns, trends, and seasonal variations. Applications such as predictive maintenance, advance demand and response, real-time fault detection and self-healing, falling conductor detection, and distribution grid load and generation prediction. Although 5G offers limited support for AI. 6G is designed to be AI native, empowering the SG to operate, reconfigure, and self-heal autonomously in real time. Table \ref{tab:newcab6g} illustrates how the new capabilities of 6G can increase the overall values of complex SG systems. Through new features such as global coverage, sensing, sustainability, interoperability, AI capabilities, and positioning. The integration of 5G and 6G technologies into the SG. It significantly enhances real-time monitoring and control. Resulting in a more efficient and resilient SG system. The integration of 6G into SG can help meet the demanding and diverse requirements of different use cases and applications. The design and optimization of 5G and 6G networks will face challenges in adjusting these communication services effectively. To achieve the target KPIs outlined in Table  \ref{tab:5g6g_capabilities} and Table \ref{tab:newcab6g}, several key enabling technologies are expected to play an important role, as discussed below.  

\begin{itemize}
\item{\bf Network Slicing:}
Network Slicing is an essential component of a 5G system.  It allows for the creation of custom logical networks designed to meet specific QoS requirements for a variety of services. Network slices can be categorized into standard network slices and non-standard network slices. Standard network slices simplify the deployment of services and ensure interoperability between devices and services, as well as among different service providers. On the other hand, non-standard slices, which are typically defined by operators, provide increased flexibility for use cases and consumers as needed. The standardized network slices include: 
\begin{itemize}
    \item eMBB: Designed to support high data rate applications, such as ultra-high-definition video streaming and augmented or virtual reality (AR/VR) \cite{b27}.
    \item URLLC: Essential for mission-critical services that demand near-instantaneous data transmission, including autonomous driving and remote surgical operations \cite{b27}.
    \item Massive Internet of Things (mIoT): Enables large-scale deployments of low-data-rate sensors and devices, making it suitable for smart cities and industrial monitoring systems \cite{b27}.
    \item Vehicle-to-Everything (V2X): Supports intelligent transportation systems by facilitating seamless connectivity between vehicles, infrastructure, and other network entities \cite{b27}.
    \item High-Performance Machine-Type Communications (HMTC): Optimized for industrial automation and robotics, requiring both high reliability and performance \cite{b27}.
    \item High Data Rate and Low Latency Communications (HDLLC): Represents a hybrid approach, combining the high throughput of eMBB with the ultra-low latency of URLLC to serve applications with demanding communication requirements \cite{b27}.
    \item Guaranteed Bit Rate Streaming Services (GBRSS): Ensure consistent and reliable throughput, making them ideal for latency-sensitive media streaming applications \cite{b27}.
\end{itemize}
6G may bring the concept of dynamic network slicing which allows the capability to adjust to varying service demands and network conditions. Dynamic network slices are dynamically created, modified, and deleted with resources allocated flexibly; they can be used to improve the performance of different energy vertical applications.
\item{\bf Service Isolation:}
SG resources can be effectively and resiliently isolated from other resources while adhering to service-level agreements (SLA). This isolation not only improves security and privacy but also enhances reliability by guaranteeing the dedicated resources necessary for ultra-high reliability in protection-related substation communications \cite{b7}.

\item{\bf Edge Computing:}
Edge Computing, or Multi-Access-Edge Computing (MEC), has the potential to significantly reduce the end-to-end latency for time-sensitive applications or traffic on the SG \cite{b29}.  A MEC platform refers to a set of fundamental functions that allows MEC applications to run on a specific virtualization infrastructure, facilitating the delivery and consumption of MEC services. In the overall architecture, user equipment (UE), Radio Access Network (RAN), and MEC, all are located nearby. There is a notable improvement in the end-to-end latency compared to a traditional 5G network. In the latter case, the RAN, 5G core, and Data Network are situated in different locations within the given area. 

\item{\bf Service Enabler Architecture Layer:}
The Service Enabler Architecture Layer (SEAL) serves as a rich functional architecture layer over the 3GPP network, empowering vertical applications such as SG communication and effectively supporting mission-critical applications \cite{b30}. It provides a flexible, secure, and scalable architecture that supports advanced applications and technologies. By offering customizable functionalities, real-time communication, and robust security, SEAL empowers SG to become more efficient, resilient, and adaptive, making it indispensable for SG energy systems.

\item{\bf Radio Air Interface:}
The 5G New Radio (NR) air interface has introduced significant complexity and flexibility compared to 4G LTE. It features scalable numerologies and various frame structures. As we move toward 6G, its air interface is anticipated to become even more advanced. It caters to extreme performance demands across a range of applications, spectrum bands, and environments. Key innovations under consideration include novel waveform designs such as Orthogonal Time Frequency Space (OTFS) and Non-Orthogonal Multiple Access (NOMA). OTFS provides robustness in high-mobility and doubly dispersive channels. NOMA facilitates massive device connectivity and efficient spectrum utilization in uncoordinated access situations \cite{b11, b25}. 

Enhancements such as a shorter TTI resulting from wider subcarrier spacing for an Orthogonal Frequency Division Multiplexing (OFDM) based air interface. Hyper Reliable and Low-Latency Communication (HRLLC)-centric Channel Quality Indicator (CQI) tables, distributed RAN architecture with high-performance edge computing, advanced scheduling (e.g., intelligent pre-configuration of radio resources), and high-performance advanced antenna techniques such as massively distributed Multiple-Input Multiple-Output (MIMO) that significantly increases Signal-to-Interference-plus-Noise Ratio (SINR) are key contributors to help achieve 0.1 ms latency and 99.9999\% reliability.

To further improve spectral efficiency and minimize latency, in-band full duplexing is being explored. This approach allows for simultaneous uplink and downlink transmissions on the same frequency band. Additionally, the integration of cutting-edge antenna technologies is crucial for the advancement of 6G. These technologies encompass RIS for programmable wireless propagation, Digital Twins for real-time network modeling, Massively Distributed MIMO and Cell-Free MIMO to ensure consistent service quality, holographic beamforming for precise spatial control, and Orbital Angular Momentum (OAM) to enhance capacity through additional degrees of freedom \cite{b23, b25}. Together, these innovations aim to create a flexible, intelligent, and reconfigurable radio interface architecture suitable for next-generation communication systems \cite{b11, b25, b31}.

\item{\bf Spectrum Sharing Mechanism:}
6G will place a strong emphasis on resource sharing, particularly in relation to spectrum, utilizing various methods. 6G, like 5G, will support traditional frequency range 1 (FR1), which operates below 7 GHz, and Frequency Range 2 (FR2), covering frequencies from 24 GHz to approximately 70 GHz. Additionally, 6G is expected to utilize the mid-band spectrum in the range 7 to 24 GHz. 6HG may support sub-THz and THz frequency bands such as 100 GHz to 300 GHz and 300 GHz to 3 THz. 

\item{\bf ISAC:}
ISAC is also known as Joint Communication and Sensing (JCAS). It is a key feature of 6G technology. ISAC uses radio signals for both communication and sensing \cite{b23, b25}. For example, a radio signal can transmit user traffic while also detecting characteristics of an object or an environment. In the emerging landscape of SG environments, the growing use of autonomous systems, including drones and UAVs, presents operational and security challenges near critical power infrastructure. Unauthorized intrusions by these UAVs or vehicles into transmission stations, substations, or overhead line passages can result in equipment damage, electrical hazards, or disruptions to service. The UAVs/Vehicles/Pedestrians detection near SG equipment undertakes the ISAC capabilities of 5G/6G systems to facilitate real-time detection and trajectory tracking of nearby entities. By utilizing base stations equipped with sensing functionalities, the system continuously monitors designated restricted zones and alerts operators when unauthorized objects such as UAVs, vehicles, or pedestrians breach a defined safety threshold \cite{b32}. Moreover, distributed sensing is expected to be an important enabler of ISAC \cite{b25, b32}.

\item{\bf AI-Native Design:}
AI/ML is becoming increasingly crucial in wireless communications, evolving from specific applications in 5G to a foundational role in 6G. In 5G and 5G advanced, AI/ML can be utilized in an implementation-specific manner as part of the Self-Organizing Network (SON) and Network Data Analytics Function (NWDAF). 6G is anticipated to be AI-native, with AI embedded from the design phase \cite{b25}. AI-native 6G will facilitate real-time decision-making, continuous learning, and optimization for enhanced performance, reliability, and user experience. SG can benefit from enhanced communication reliability, real-time optimization, advanced sensing, and robust security. Furthermore, the SG can become more autonomous, adaptive, and efficient. AI-native 6G can help the SG meet the demand for smart energy systems, integrate renewable resources seamlessly, and ensure a sustainable energy future.

\item{\bf Sustainability:}
The increased complexity of 6G compared to 5G may result in higher energy consumption unless sustainability-specific actions are taken. It is important to define performance metrics that measure sustainability in both the network and the device. Technologies for a green radio network utilizing frequency, time, spatial, and power domains are essential. While 6G aims to achieve substantial energy efficiency improvements over 5G, the introduction of more complex architectures such as AI-native air interfaces, large-scale MIMO, holographic metasurfaces, and edge intelligence, tend to increase the power consumption of communication infrastructure. In the context of SGs, these advancements facilitate system-level savings by reducing inefficiencies. The Next G Alliance has identified several sustainability features for 6G \cite{b31} and carried out a study on sustainable AI in telecom \cite{b47}. 6G is expected to have both network-side power saving methods and device-side power saving methods. NZE communication is also being mentioned for 6G. 6G enhances the alignment between demand and supply, improves forecasting accuracy, and enables precise control over DERs and loads. This creates a trade-off between the heightened local energy usage associated with 6G and the potential for significant reductions in overall grid energy consumption and emissions. Therefore, a comprehensive sustainability assessment that spans both communication and power domains is essential.

\item{\bf Digital Twin:}
A Digital Twin is defined as a real-time digital replica of a physical system, object, or process that remains synchronized with its physical counterpart through continuous data exchange \cite{b23, b25}. Unlike conventional digital shadows, which operate in a one-way manner, It facilitate two-way interaction, allowing for monitoring, simulation, prediction, and optimization. In the realm of 5G and 6G-enabled SG. This empowers operators to simulate energy flows, anticipate potential failures, optimize resource allocation, and implement proactive control strategies without disrupting physical operations. 

For example, a transformer equipped with temperature, vibration, dissolved gas, and load sensors. These measurements are transmitted via 5G/6G networks to an edge- or cloud-hosted digital twin that continuously updates state and lifetime models. The twin can detect early signs of aging or faults, prompting tap changer adjustments, load redistribution, or maintenance actions. Low latency is critical for protection decisions in response to over-temperature conditions or abnormal vibrations.

The advent of 6G technologies, such as integrated AI, ultra-low-latency communication, and edge computing, significantly enhances the functionality of digital twins by enabling faster, autonomous, and context-aware responses to dynamic grid conditions. This aligns with critical 6G objectives, including sustainability, resilience, and trustworthiness, and are regarded as foundational tools for the effective management of smart infrastructure in energy systems, urban settings, and industrial automation \cite{b33}.

\item {\bf Practical Implementation Considerations:} While holographic MIMO, RIS, and cell-free architectures promise enhanced coverage and spectral efficiency for 6G SGs. The practical implementations encounter substantial challenges from hardware impairments. These include phase noise, finite-resolution phase control, transceiver non-linearities, and element mutual coupling. Recent research on stacked intelligent metasurfaces \cite{b44}, holographic metasurface beamforming for multi-altitude Low-Earth-Orbit (LEO) satellite networks \cite{b45}, and reconfigurable holographic surfaces under hardware impairments \cite{b46} quantifies the performance degradation relative to ideal theoretical models. Consequently, SG 6G deployments must account for these factors when dimensioning reliability and latency margins to ensure mission-critical operation.
\end{itemize}

\section{Challenges and Future Directions}
\label{sec:challenges}
The various key enabling technologies/mechanisms discussed in Section \ref{sec:capab} help in the design and optimization of 5G and 6G networks to meet the service requirements of various SG use cases. However, several open challenges need to be addressed, including efficient network slicing design and management, scheduling and resource allocation, low latency improvement, intelligent edge computing, spectrum management, and newly introduced capabilities of 6G to make a 5G/6G-based SG a reality.

\begin{itemize}
\item{\bf Efficient Network Slicing Design and Management:}
Efficient network slice design and management are important for empowering SG powered by 5G and 6G networks. It ensures seamless communication, scalability, and reliability for a variety of grid use cases \cite{b38}. As discussed in Section \ref{sec:capab}, network slicing plays a significant role in this context. The SG incorporates a vast number of smart meters, sensors, and IoT devices tailored to specific use cases, resulting in diverse service requirements as outlined in Table \ref{tab:sgucservicerequirements}. To address these requirements, multiple network slices will need to be created, configured, activated/deactivated, and terminated. Ensuring each slice meets the target QoS requirements presents several challenges such as dynamic slice adjustment, multidomain and multioperator support, high reliable and secure slicing \cite{b37}. These include scalability to support millions of IoT devices, ensure ultra-low latency for critical use cases such as real-time fault detection and self-healing, advance demand and response, balance energy efficiency with performance and reliability, and achieve interoperability among various SG applications. 

For sudden grid events, such as line faults causing monitoring traffic surges, the following approaches can be utilize: (1) Pre-provisioned URLLC "protection slices" with resource reservation to guarantee fault detection and protection QoS. (2) Dynamic elastic monitoring slices (e.g., eMBB) that auto-scale via local closed-loop Network Slice Management Function (NSMF)/Network Slice Subnet Management Function (NSSMF) orchestration, preempting only non-critical metering traffic while preserving protection slice isolation. (3) Graceful post-recovery deletion with resource reclamation to the elastic pool. These strategies maintain strict end-to-end SLA compliance during surges \cite{b39, b40}.
To address these challenges, research should concentrate on the development of AI-based Network Slice management such as autonomous network slice creation, configuration, activation/deactivation, slice isolation, slice coordination, and energy-efficient design \cite{b37}. By developing innovative network slice design with proactive management, 5G and 6G can support a variety of SG applications, ensuring an efficient and sustainable energy system \cite{b21, b37, b38}.

\item {\bf End-to-End Security and Zero-Trust:} Efficient network slice management is crucial for ensuring QoS and resource availability. However, the shared physical infrastructure of 5G and 6G networks introduces complex cyber threats. To secure SG, a comprehensive end-to-end security architecture is essential. The decentralized deployment of DERs significantly expands the grid's attack surface \cite{b6}, underscoring the need for stringent zero-trust principles. This means that continuous authentication and authorization must be implemented at the grid edge rather than relying solely on traditional perimeter defenses. Additionally, strong cryptographic slice isolation is vital for protecting against cross-slice attacks. It is imperative to prevent vulnerabilities exploited in a large IoT slice, such as residential smart meters, from being leveraged to infiltrate mission-critical URLLC slices, like high speed differential protection. Future research should prioritize AI-driven threat detection to dynamically enforce zero-trust policies and provide cross-slice protection in diverse SG environments.

\item{\bf Standardizing the 5G/6G KPIs:}
KPIs are very important to monitor the performance of SG use cases. In Table \ref{tab:sgucservicerequirements}, several KPIs for SG use cases have not yet been investigated or are not well defined, such as availability, reliability, data rate, latency, jitter, and message size for SG use cases. New capabilities defined in Table \ref{tab:newcab6g}, such as AI capabilities, sensing, interoperability, and positioning, can be researched for SG use cases.  Further, research can be performed to specify the values for these KPIs in the context of SG use cases. For these appearing use cases, we recommend that future work determine these values by: (i) decomposing such use cases into well-specified primitive functions, (ii) collecting empirical measurements from field deployments (e.g., PMUs, smart meters, and environmental sensors), (iii) performing detailed simulations, and (iv) aligning the results with emerging standards for the SG.

\item{\bf Scheduling and Resource Allocation:}
Scheduling and resource allocation are very important for the efficient operation of 5G/6G-based SG. As they ensure that critical required resources for SG use cases such as real-time fault detection and self healing, automated load balancing, and DERs coordination are prioritized and managed seamlessly. The need for effective scheduling and resource allocation strategies, to meet diverse demands includes managing time-sensitive communication, bandwidth requirements, and energy efficiency. By utilizing techniques such as AI-driven dynamic scheduling, intelligent edge computing, dynamic network slicing, and predictive resource allocation, SG systems can ensure reliability, scalability, and energy efficiency. Additionally, research should be focused on managing challenges such as inter-slice resource sharing, energy optimization, and scalability to support the growing complexity of SG systems \cite{b22, b37, b38}.

\item{\bf Intelligent Edge Computing:}
Intelligent edge computing is an essential technology for 5G/6G enabled SG use cases such as real-time fault detection and self-healing, advance demand and response, and real-time predictive analytics. By processing data locally and leveraging AI, intelligent edge computing manages latency sensitive tasks, improves scalability, and enhances grid reliability \cite{b4, b24, b38}. As the SG evolves, further research into energy-efficient, decentralized computing architectures, faster data processing, and secure edge solutions will play a crucial role in achieving resilient and sustainable energy systems \cite{b22, b26}.

\item {\bf Cohesive AI Orchestration Frameworks:} AI/ML is being applied more widely across different grid domains, ranging from device-level inference for detecting falling conductors to network-level optimization through O-RAN. However, a universally standardized, layered AI architecture has not yet been fully established in real-world applications. Future efforts should concentrate on creating unified frameworks that integrate these fragmented AI capabilities seamlessly across device, edge, and cloud environments, while meeting the strict URLLC requirements needed for mission-critical grid protection.

\item{\bf Smart Spectrum Management:}
Smart spectrum management (SSM) is a critical mechanism for reliable and efficient communications in 5G/6G based SG systems. As SG use cases rely on faster data speeds, reliable coverage, massively connected devices, and ultra-low latency. SSM ensures optimal use of available frequency bands. The main objective of SSM is to dynamically manage spectrum resources to meet the unique requirements of SG use cases . While minimizing interference, maximizing bandwidth utilization, and ensuring reliability. Key techniques such as dynamic spectrum allocation, cognitive radio networks, AI-driven optimization, and custom network slicing lead to adaptive and efficient use of spectrum resources. Managing challenges such as limited spectrum availability, spectrum efficiency, interference, and regulatory limitations. Through innovative research and technologies will be essential for upcoming advanced SG systems. By adopting these strategies, SG can achieve seamless and low latency communication to support SG systems \cite{b26, b36}.

\item{\bf O-RAN Integration in SG:}
O-RAN is an innovative way of designing and deploying a RAN. O-RAN decouples hardware and software features in the RAN, disaggregates it into modular components, and facilitates openness, intelligence, modularity, and flexibility \cite{b38}. For SG use cases, O-RAN provides flexibility and AI-based controllers to manage the RAN efficiently. Despite challenges related to interoperability, security, and latency, O-RAN's open, modular, and AI-based architecture makes it an effective RAN architecture for SG. The Near-Real-Time RAN Intelligent Controller (Near-RT RIC) operates within 10 ms–1 s, hosting near-real-time applications (xApps) that dynamically adjust radio resources and slicing parameters to support, such as smart distributed voltage control, real-time fault detection and self-healing, and slice isolation. In contrast, the Non-Real-Time RAN Intelligent Controller (Non-RT RIC) functions beyond 1 s, hosting non-real-time applications (rApps) focused on metering analytics, policy optimization, and outage prediction for SG operations \cite{b41}.

For example, an SG application (e.g., Smart Distributed Voltage Control) analyzes raw electrical telemetry measurements and securely communicates the resulting telecommunications service intents to the network's Service Management and Orchestration (SMO) framework. In response, the Non-RT RIC issues application-aware policy directives via the A1 interface to the Near-RT RIC, which in turn guides a dedicated RAN-optimization xApp tailored for voltage-control traffic. The xApp then issues E2SM control commands via the E2 interface to the O-DU to dynamically adjust MAC-layer scheduling, influencing the utilization of 5G/6G physical resource blocks, guaranteeing deterministic latency for the critical utility data flows. Conversely, an outage-prediction rApp in the Non-RT RIC analyzes historical load and weather data ingested as Enrichment Information via the SMO to proactively reconfigure network slice policies, preventing communication bottlenecks and enhancing grid resilience before extreme weather events.

Further research and development may unlock the full potential of O-RAN for building resilient, scalable, and efficient SG systems \cite{b38}. For use cases that can tolerate delays exceeding 10 ms, a Near-RT RIC with suitable xApps is adequate. Conversely, applications requiring latencies of less than 10 ms necessitate a customized O-RAN architecture that utilizes distributed applications (dApps) or micro applications (µApps) \cite{b41}. In one possible implementation, dApps or µApps can be implemented in a distributed manner such as in O-RAN-Distributed Unit (O-DU) or O-RAN Radio Unit(O-RU) depending on the use case. This implementation approach avoids reliance on RICs that are external to O-DU/O-RU. Protocols such as synchronous Ethernet, IEEE 1588 Precision Time Protocol (PTP), and Time Sensitive Networking (TSN) 802.1Qbv help ensure deterministic delay in virtualized O-RAN deployments. Together, these features allow O-RAN to support diverse applications with strict latency requirements.
\end{itemize}

\section{Summary}
\label{sec:smmry}
We have presented a comprehensive overview of the SG architecture, its essential features, and the underlying reasons for its increasing popularity. This study included a comparison between wired and wireless networks, highlighting the advantages of cellular technologies. Furthermore, we identified emerging SG use cases along with their stringent service requirements, which encompass availability, reliability, latency, data rate, connection density, jitter, and time synchronization accuracy.

We also conducted a comparative evaluation of the advanced capabilities of 5G and 6G networks, highlighting their potential to transform the SG ecosystem. Key enablers identified include Network Slicing, Edge Computing, SSM, SEAL, Digital Twin, O-RAN, PLS, ISAC, and AI-native design. These advances tackle critical challenges related to scalability, reliability, security, sustainability, and efficiency, ensuring that 5G and 6G networks can effectively support the complex demands of emerging SG.

Finally, we discussed several open challenges that must be addressed to fully realize 5G/6G-enabled SG systems. In conclusion, the integration of 5G and 6G networks with the SG represents a pivotal advancement in the pursuit of sustainable and resilient energy systems. As these networks continue to evolve, their collaboration with the SG will facilitate innovative services, improve decentralized energy management, and support the global shift toward smarter, cleaner, and more interconnected energy ecosystems. The ongoing research and collaboration between the energy and telecommunications sectors will be crucial in addressing future challenges and achieving the full potential of this convergence.

\section*{Acknowledgment}
This work was supported by Secure Telecommunication ARchitecture for Trusted and Resilient Electric Communications (STAR TREC) project, funded by the Electric Power Research Institute (EPRI) under Grant No. 10018395, and SEcure Control for REnewable DERs in Power Grid (SECURED-GRID) project under Grant No. W911NF2310211, sponsored by Army Contracting Command.

\newpage

\begin{IEEEbiography}[{\includegraphics[width=1in,height=1.25in,clip,keepaspectratio]{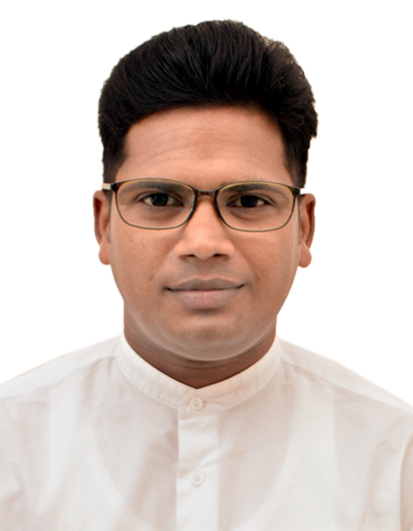}}]{Manoj Kumar} received his Master of Science degree in Computer Science from the University of Allahabad, India, in 2014, and the Post Graduate Diploma in Advanced Computing from the Centre for Development of Advanced Computing (C-DAC), Hyderabad, India, in 2017. He is currently pursuing a Ph.D. degree in Electrical Engineering at Virginia Tech, Blacksburg, USA. Before joining the PhD program, he worked at Rakuten Mobile Inc., Tokyo, Japan (Jan-2020 - Jul-2023) and WiSig Networks, Hyderabad, India (Nov-2017 - Nov-2019). His research interests include low-latency communications, energy-efficient 5G/6G systems and beyond, NTN, O-RAN, AI/ML, and the Smart Grid.
\end{IEEEbiography}
\begin{IEEEbiography}[{\includegraphics[width=1.09in,height=1.25in,clip]{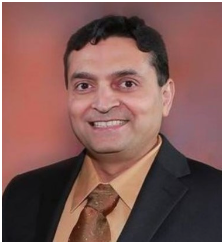}}]{Dr. Nishith Tripathi} is a research associate professor at Virginia Tech. Dr. Tripathi has 24 years of hands-on industry experience and is an expert on various aspects of commercial 3G, 4G, and 5G wireless networks including design, operations, testing, and optimization. At Virginia Tech, he has led several sponsored research projects on 5G, 5G-Advanvced, and 6G in the areas such as O-RAN testbeds, SpaceNet testbed, O-RAN xApps, O-RAN testing, enhanced security for 5G and 6G, NTN, V2X communications, geofencing, positioning, UAV/UAS, and smart warehouses. He has co-authored five books including the world’s first multimedia book on 5G, a comprehensive textbook on cellular communications, a pioneering monograph on the RRM using AI, a book on O-RAN fundamentals, and a book on 3GPP-based NTN. He has made more than thirty contributions to the development of the 3GPP 5G specifications. He is an academic engagement lead at the Next G Alliance and has contributed to several NGA white papers. He leads a 5G/FutureG Working Group at the NSC to develop 3GPP contributions on 5G-Advanced and 6G. As a wireless industry expert, Dr. Tripathi has contributed to organizations such as FCC, CTIA, GSMA, NGA, NSC, Scientific American, FTC, EE Times University, and CNN. He is a founder of Aum Research and Consulting (ARC) that provides research and consulting services.
\end{IEEEbiography}
\begin{IEEEbiography}[{\includegraphics[width=1.09in,height=1.25in,clip, keepaspectratio]{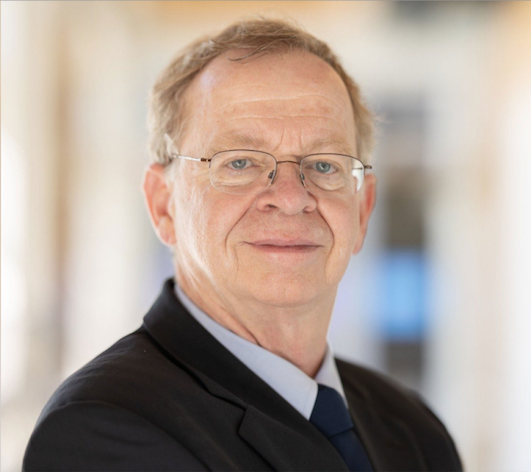}}]{Professor Jeffrey Reed} is the Willis G. Worcester Professor of ECE. Professor Reed’s research interests are wireless communications, wireless security, cognitive radio, software radio, telecommunications policy, and spectrum access. Reed has co-authored more than 500 articles and books. In addition, Reed co-founded several commercial companies, including Federated Wireless, which commercializes spectrum sharing; PFP Cybersecurity, which provides security solutions for IoT devices; and Cirrus360, which produces tools for O-RAN development and private networks. Reed is the Founding Director of Wireless@Virginia Tech, a university research center, and co-founder of Virginia Tech’s Hume Center for National Security and Technology, where he served as the Interim Director. He also served as the Interim Director of the Commonwealth Cyber Initiative and is currently its CTO. Dr. Reed is a Fellow of the IEEE for contributions to software radio and communications signal processing and leadership in engineering education.  In 2025, Dr. Reed was elected to the National Academy of Inventors.
\end{IEEEbiography}

\end{document}